\input harvmac
\overfullrule=0pt
\parindent 25pt
\tolerance=10000

\input epsf

\newcount\figno
\figno=0
\def\fig#1#2#3{
\par\begingroup\parindent=0pt\leftskip=1cm\rightskip=1cm\parindent=0pt
\baselineskip=11pt
\global\advance\figno by 1
\midinsert
\epsfxsize=#3
\centerline{\epsfbox{#2}}
\vskip 12pt
{\bf Fig.\ \the\figno: } #1\par
\endinsert\endgroup\par
}
\def\figlabel#1{\xdef#1{\the\figno}}
\def\encadremath#1{\vbox{\hrule\hbox{\vrule\kern8pt\vbox{\kern8pt
\hbox{$\displaystyle #1$}\kern8pt}
\kern8pt\vrule}\hrule}}

\def\half{{\textstyle{1\over2}}}

\def\apm{{\alpha^{\prime}}}

\def\half{{1\over 2}}
 
 \def\m{{\mu}}
 \def\w{{\omega}}       
 \def\n{{\nu}}

 \def\rh{\rho}
 \def\t{{\theta}}
 \def\a{{\alpha}}
 
 \def\frac#1#2{{#1\over #2}}

 \def\s{{\sigma}}
 
 \def\b{{\beta}}

 \def\CO{{\cal O}}
 
 \def\Ph{{\Phi }}
 \def\L{{\Lambda}}
 
 \def\p{\partial}

 \def\apm{{\alpha'}}
 \def\r{\rightarrow}
 \def\Re{{\rm Re}}


\lref\DixonIZ{
L.~J.~Dixon and J.~A.~Harvey,
``String Theories In Ten-Dimensions Without Space-Time Supersymmetry,''
Nucl.\ Phys.\ B {\bf 274}, 93 (1986).
}

\lref\BardakciGV{
K.~Bardakci,
``Dual Models And Spontaneous Symmetry Breaking,''
Nucl.\ Phys.\ B {\bf 68}, 331 (1974).
}

\lref\BardakciAN{
K.~Bardakci,
``Spontaneous Symmetry Breakdown In The Standard Dual String Model,''
Nucl.\ Phys.\ B {\bf 133}, 297 (1978).
}

\lref\BardakciVS{
K.~Bardakci and M.~B.~Halpern,
``Explicit Spontaneous Breakdown In A Dual Model,''
Phys.\ Rev.\ D {\bf 10}, 4230 (1974).
}

\lref\BardakciUX{
K.~Bardakci and M.~B.~Halpern,
``Explicit Spontaneous Breakdown In A Dual Model. 2. N Point Functions,''
Nucl.\ Phys.\ B {\bf 96}, 285 (1975).
}

\lref\RussoNA{
J.~G.~Russo and A.~A.~Tseytlin,
``Supersymmetric fluxbrane intersections and closed string tachyons,''
arXiv:hep-th/0110107.
}

\lref\SchmidhuberBV{
C.~Schmidhuber and A.~A.~Tseytlin,
``On string cosmology and the RG flow in 2-d field theory,''
Nucl.\ Phys.\ B {\bf 426}, 187 (1994)
[arXiv:hep-th/9404180].
}

\lref\BergmanRW{
A.~Bergman, K.~Dasgupta, O.~J.~Ganor, J.~L.~Karczmarek and G.~Rajesh,
``Nonlocal field theories and their gravity duals,''
arXiv:hep-th/0103090.
}

\lref\PolchinskiRQ{
J.~Polchinski,
``String Theory. Vol. 1: An Introduction To The Bosonic String,''
{\it  Cambridge, UK: Univ. Pr. (1998) 402 p}.
}

\lref\RussoTF{
J.~G.~Russo and A.~A.~Tseytlin,
 ``Magnetic backgrounds and tachyonic instabilities in closed superstring  
theory and M-theory,''
Nucl.\ Phys.\ B {\bf 611}, 93 (2001)
[arXiv:hep-th/0104238].
}
\lref\TakayanagiJJ{
T.~Takayanagi and T.~Uesugi,
``Orbifolds as Melvin geometry,''
arXiv:hep-th/0110099.
}

\lref\DudasUX{
E.~Dudas and J.~Mourad,
``D-branes in string theory Melvin backgrounds,''
arXiv:hep-th/0110186.
}

\lref\TakayanagiAJ{
T.~Takayanagi and T.~Uesugi,
``D-branes in Melvin background,''
arXiv:hep-th/0110200.
}

\lref\WittenQJ{
E.~Witten,
``Anti-de Sitter space and holography,''
Adv.\ Theor.\ Math.\ Phys.\  {\bf 2}, 253 (1998)
[arXiv:hep-th/9802150].
}

\lref\SaffinKY{
P.~M.~Saffin,
Phys.\ Rev.\ D {\bf 64}, 024014 (2001)
[arXiv:gr-qc/0104014].
}

\lref\DabholkarIF{
A.~Dabholkar,
``Tachyon condensation and black hole entropy,''
arXiv:hep-th/0111004.
}

\lref\DabholkarGZ{
A.~Dabholkar,
``On condensation of closed-string tachyons,''
arXiv:hep-th/0109019.
}

\lref\vafaatish{
A.~Dabholkar and C.~Vafa,
``tt* Geometry and closed string tachyon potential,''
arXiv:hep-th/0111155.
}

\lref\chicago{ J.~A.~Harvey, D.~Kutasov, E.~J.~Martinec, G.~Moore,
``Localized tachyons and RG flows'', arXiv:hep-th/0111154.
}

\lref\MaldacenaRE{
J.~Maldacena,
``The large $N$ limit of superconformal field theories and supergravity,''
Adv.\ Theor.\ Math.\ Phys.\  {\bf 2}, 231 (1998)
[arXiv:hep-th/9711200].
}

\lref\WittenYC{
E.~Witten,
``Phases of N = 2 theories in two dimensions,''
Nucl.\ Phys.\ B {\bf 403}, 159 (1993)
[arXiv:hep-th/9301042].
}

\lref\PolyakovJU{
A.~M.~Polyakov,
``The wall of the cave,''
Int.\ J.\ Mod.\ Phys.\ A {\bf 14}, 645 (1999)
[arXiv:hep-th/9809057].
}

\lref\CostaIF{
M.~S.~Costa, C.~A.~Herdeiro and L.~Cornalba,
``Flux-branes and the dielectric effect in string theory,''
arXiv:hep-th/0105023.
}

\lref\DixonQV{
L.~J.~Dixon, D.~Friedan, E.~J.~Martinec and S.~H.~Shenker,
``The Conformal Field Theory Of Orbifolds,''
Nucl.\ Phys.\ B {\bf 282}, 13 (1987).
} 

\lref\SuyamaGD{
T.~Suyama,
``Properties of string theory on Kaluza-Klein Melvin background,''
arXiv:hep-th/0110077.
}

\lref\ChenNR{
C.~M.~Chen, D.~V.~Gal'tsov and P.~M.~Saffin,
``Supergravity fluxbranes in various dimensions,''
arXiv:hep-th/0110164.
}

\lref\RussoTF{
J.~G.~Russo and A.~A.~Tseytlin,
 ``Magnetic backgrounds and tachyonic instabilities in closed superstring  theory and M-theory,''
Nucl.\ Phys.\ B {\bf 611}, 93 (2001)
[arXiv:hep-th/0104238].
}

\lref\EmparanGM{
R.~Emparan and M.~Gutperle,
``From p-branes to fluxbranes and back,''
arXiv:hep-th/0111177.
}

\lref\GavaJT{
E.~Gava, K.~S.~Narain and M.~H.~Sarmadi,
``On the bound states of p- and (p+2)-branes,''
Nucl.\ Phys.\ B {\bf 504}, 214 (1997)
[hep-th/9704006].
}

\lref\BuscherQJ{
T.~H.~Buscher,
``Path Integral Derivation Of Quantum Duality In Nonlinear Sigma Models,''
Phys.\ Lett.\ B {\bf 201}, 466 (1988).
}

\lref\GreenUM{
M.~B.~Green and M.~Gutperle,
``Light-cone supersymmetry and D-branes,''
Nucl.\ Phys.\ B {\bf 476}, 484 (1996)
[arXiv:hep-th/9604091].
}

\lref\TseytlinZV{
A.~A.~Tseytlin,
``Closed superstrings in magnetic flux tube background,''
Nucl.\ Phys.\ Proc.\ Suppl.\  {\bf 49}, 338 (1996)
[arXiv:hep-th/9510041].
}

\lref\TseytlinEI{
A.~A.~Tseytlin,
``Melvin solution in string theory,''
Phys.\ Lett.\ B {\bf 346}, 55 (1995)
[arXiv:hep-th/9411198].
}

\lref\BanksCH{
T.~Banks and L.~Susskind,
``Brane - Antibrane Forces,''
arXiv:hep-th/9511194.
}

\lref\TaylorIK{
W.~I.~Taylor,
``D-brane field theory on compact spaces,''
Phys.\ Lett.\ B {\bf 394}, 283 (1997)
[arXiv:hep-th/9611042].
}

\lref\gibbonsa{
G.~W.~Gibbons and K.~Maeda,
``Black Holes And Membranes In Higher Dimensional Theories With Dilaton
Fields,'' 
Nucl.\ Phys.\ B {\bf 298}, 741 (1988).
}

\lref\ScherkTA{
J.~Scherk and J.~H.~Schwarz,
``Spontaneous Breaking Of Supersymmetry Through Dimensional Reduction,''
Phys.\ Lett.\ B {\bf 82}, 60 (1979).
}

\lref\greglaf{
R.~Gregory and R.~Laflamme,
``Black Strings And P-Branes Are Unstable,''
Phys.\ Rev.\ Lett.\  {\bf 70}, 2837 (1993)
[arXiv:hep-th/9301052].
}

\lref\HorowitzCZ{
G.~T.~Horowitz and K.~Maeda,
``Fate of the black string instability,''
Phys.\ Rev.\ Lett.\  {\bf 87}, 131301 (2001)
[arXiv:hep-th/0105111].
}

\lref\SaffinKY{
P.~M.~Saffin,
``Gravitating fluxbranes,''
Phys.\ Rev.\ D {\bf 64}, 024014 (2001)
[arXiv:gr-qc/0104014].
}

\lref\CostaIF{
M.~S.~Costa, C.~A.~Herdeiro and L.~Cornalba,
 ``Flux-branes and the dielectric effect in string theory,''
arXiv:hep-th/0105023.
}

\lref\GubserAC{
S.~S.~Gubser,
``On non-uniform black branes,''
arXiv:hep-th/0110193.
}

\lref\witbub{E.~Witten,
 ``Instability Of The Kaluza-Klein Vacuum'',
Nucl.\ Phys.\  {\bf B195} (1982) 481.
}

\lref\hort{F.~Dowker, J.~P.~Gauntlett, G.~W.~Gibbons and G.~T.~Horowitz,
``Nucleation of $P$-Branes and Fundamental Strings'',
Phys.\ Rev.\  {\bf D53} (1996) 7115
[hep-th/9512154].}

\lref\garstr{D.~Garfinkle and A.~Strominger,
``Semiclassical Wheeler wormhole production'',
Phys.\ Lett.\  {\bf B256} (1991) 146.
}

\lref\costa{
M.~S.~Costa and M.~Gutperle,
``The Kaluza-Klein Melvin solution in M-theory,''
JHEP {\bf 0103}, 027 (2001)
[arXiv:hep-th/0012072].
}

\lref\berg{
O.~Bergman and M.~R.~Gaberdiel,
``Dualities of type 0 strings,''
JHEP {\bf 9907}, 022 (1999)
[arXiv:hep-th/9906055].
}

\lref\tse{
J.~G.~Russo and A.~A.~Tseytlin,
``Magnetic flux tube models in superstring theory,''
Nucl.\ Phys.\ B {\bf 461}, 131 (1996)
[arXiv:hep-th/9508068].
}

\lref\callan{
C.~G.~Callan and J.~M.~Maldacena,
``Brane dynamics from the Born-Infeld action,''
Nucl.\ Phys.\ B {\bf 513}, 198 (1998)
[arXiv:hep-th/9708147].
}

\lref\horava{
M.~Fabinger and P.~Horava,
``Casimir effect between world-branes in heterotic M-theory,''
Nucl.\ Phys.\ B {\bf 580}, 243 (2000)
[arXiv:hep-th/0002073].
}

\lref\melvin{
M.~A.~Melvin, ``Pure Magnetic and Electric Geons'',
Phys. Lett. {\bf 8} (1964) 65.
}

\lref\dowkerb{
F.~Dowker, J.~P.~Gauntlett, G.~W.~Gibbons and G.~T.~Horowitz,
``The Decay of magnetic fields in Kaluza-Klein theory,''
Phys.\ Rev.\ D {\bf 52} (1995) 6929
[hep-th/9507143].
}

\lref\sena{
A.~Sen,
``Stable non-BPS states in string theory,''
JHEP {\bf 9806}, 007 (1998)
[arXiv:hep-th/9803194].
}

\lref\aps{
A.~Adams, J.~Polchinski and E.~Silverstein,
``Don't panic! Closed string tachyons in ALE space-times,''
arXiv:hep-th/0108075.
}

\lref\seibwit{
N.~Seiberg and E.~Witten,
``Spin Structures In String Theory,''
Nucl.\ Phys.\ B {\bf 276}, 272 (1986).
}

\lref\gibbonsa{
G.~W.~Gibbons and K.~Maeda,
``Black Holes And Membranes In Higher Dimensional Theories With Dilaton
Fields,'' 
Nucl.\ Phys.\ B {\bf 298}, 741 (1988).
}

\lref\andya{
M.~Gutperle and A.~Strominger,
``Fluxbranes in string theory,''
JHEP {\bf 0106}, 035 (2001)
[arXiv:hep-th/0104136].
}

\lref\rohm{
R.~Rohm,
``Spontaneous Supersymmetry Breaking In Supersymmetric String Theories,''
Nucl.\ Phys.\ B {\bf 237}, 553 (1984).
}

\lref\atickwitten{
J.~J.~Atick and E.~Witten,
``The Hagedorn Transition And The Number Of Degrees Of Freedom Of String Theory,''
Nucl.\ Phys.\ B {\bf 310}, 291 (1988).
}

\lref\berga{
O.~Bergman and M.~R.~Gaberdiel,
``Dualities of type 0 strings,''
JHEP {\bf 9907}, 022 (1999)
[arXiv:hep-th/9906055].
}

\lref\vafa{
C.~Vafa,
``Mirror symmetry and closed string tachyon condensation,''
arXiv:hep-th/0111051.
}

\lref\mirrorsym{
K.~Hori and C.~Vafa,
``Mirror symmetry,''
arXiv:hep-th/0002222.
}

\lref\kentaro{
K.~Hori and A.~Kapustin,
``Duality of the fermionic 2d black hole and N = 2 Liouville theory as  mirror symmetry,''
JHEP {\bf 0108}, 045 (2001)
[arXiv:hep-th/0104202].
}

\lref\BlumGW{
J.~D.~Blum and K.~R.~Dienes,
``Strong/weak coupling duality relations for non-supersymmetric string
theories,''
Nucl.\ Phys.\ B {\bf 516}, 83 (1998)
[arXiv:hep-th/9707160].
}

\lref\BlumCS{
J.~D.~Blum and K.~R.~Dienes,
``Duality without supersymmetry: The case of the SO(16) x SO(16) string,''
Phys.\ Lett.\ B {\bf 414}, 260 (1997)
[arXiv:hep-th/9707148].
}

\lref\Suyamao{
T.~Suyama,
``Properties of string theory on Kaluza-Klein Melvin background,''
arXiv:hep-th/0110077.
}

\lref\Suyamat{
T.~Suyama,
``Melvin background in heterotic theories,''
arXiv:hep-th/0107116.
}

\lref\Suyamath{
T.~Suyama,
``Closed string tachyons in non-supersymmetric heterotic theories,''
JHEP {\bf 0108}, 037 (2001)
[arXiv:hep-th/0106079].
}

\lref\DasDS{
S.~R.~Das, S.~Naik and S.~R.~Wadia,
``Quantization Of The Liouville Mode And String Theory,''
Mod.\ Phys.\ Lett.\ A {\bf 4}, 1033 (1989).
}

\lref\PolchinskiFN{
J.~Polchinski,
``A Two-Dimensional Model For Quantum Gravity,''
Nucl.\ Phys.\ B {\bf 324}, 123 (1989).
}

\lref\DasDA{
S.~R.~Das, A.~Dhar and S.~R.~Wadia,
``Critical Behavior In Two-Dimensional Quantum Gravity And Equations Of
Motion Of The String,''
Mod.\ Phys.\ Lett.\ A {\bf 5}, 799 (1990).
}

\lref\BanksQE{
T.~Banks and J.~Lykken,
``String Theory And Two-Dimensional Quantum Gravity,''
Nucl.\ Phys.\ B {\bf 331}, 173 (1990).
}

\lref\piljin{
Y.~Michishita, P. Yi, ``D-brane Probe and Closed String,'' hep-th/0111199.
}

\baselineskip 18pt plus 2pt minus 2pt

\Title{\vbox{\baselineskip12pt \hbox{hep-th/0111212}\hbox{HUTP-01/A059}
  }}
{\vbox{\centerline{Closed String Tachyon Condensation on Twisted Circles}}}
\centerline{Justin R. David${}^1$, Michael Gutperle${}^2$, Matthew Headrick${}^2$, and Shiraz
  Minwalla${}^2$ } 
\medskip\centerline{${}^1$Department of Physics}
\centerline{University of California, Santa Barbara, CA 93106}

\medskip\centerline{${}^2$ Jefferson Physical Laboratory}
\centerline{Harvard University}
\centerline{Cambridge, MA 02138}

\vskip .1in \centerline{\bf Abstract}

We study IIA/B string theory compactified on twisted 
circles. These models possess closed string tachyons and reduce to  
type 0B/A theory in a special limit. Using methods of gauged linear sigma 
models and mirror symmetry we construct a conformal field theory 
which interpolates between these models and flat space
via an auxiliary Liouville direction. Interpreting motion in the
Liouville direction as renormalization group flow, we argue that
the end point of tachyon condensation in all these models (including 
0B/A theory) is supersymmetric type II theory. We also find a 
zero-slope limit of these models which is best
described in a T-dual picture as a type II NS-NS fluxbrane.
In this limit tachyon condensation is an interesting and 
well posed problem in supergravity. We explicitly determine the tachyon 
as a fluctuation of supergravity fields, and perform a rudimentary 
numerical analysis of the relevant flows.

\noblackbox

\Date{November 2001}

\listtoc
\writetoc

\newsec{Introduction}

Worldsheet conformal field theories are solutions to the equations of
motion of weakly coupled string theory.
Conformal field theories that possess no relevant operators 
are stable solutions; small excitations about the corresponding 
spacetime background have positive energy.  
On the other hand, conformal field theories 
possessing relevant operators 
are tachyonic or 
unstable solutions of string theory.\foot{For early work on tachyon
condensation in string theories see 
\refs{\BardakciGV,\BardakciAN,\BardakciVS,\BardakciUX}.} 
It is of great interest
to follow the tachyon condensation processes about such backgrounds, 
and to determine their endpoints.

There has been tremendous progress in the understanding of open string
tachyon condensation, following the work of Sen \sena. 
Open string tachyons
appear in non-supersymmetric configurations of branes, and the dynamics
of tachyon condensation has been studied using a variety of techniques,
among them string field theory, conformal field theory, and noncommutative
geometry. The picture that emerges from this study is simple and 
physically satisfying. For instance, the tachyonic mode that appears 
when a D-brane is brought next to an antibrane is a 
signal of the tendency of these objects to annihilate. 
The endpoint of this tachyon condensation is simply empty space. 

An open string tachyon is a boundary field on the worldsheet of the
string. In the $g_{\rm s} \to 0$ limit, open string tachyon 
condensation modifies  boundary conditions on the string worldsheet, 
but leaves the bulk conformal field theory untouched. In contrast,  
a zero-momentum closed string tachyon is a relevant deformation of the 
string sigma model. Consequently, closed string tachyon condensation 
modifies the bulk CFT and changes the spacetime on which the 
string propagates. Thus the dynamics of closed string tachyon
condensation may be expected to be more
interesting, but more complicated, than in the open string case.

Nonetheless Adams, Polchinski, and Silverstein \aps\ have recently 
argued for a pattern of closed string tachyon 
condensation about certain non-supersymmetric orbifolds that
qualitatively resembles that of open string tachyon condensation. 
They argue, for example, that  $C/ Z_n$ orbifolds 
of type II theory are finite energy, unstable excitations 
of type II theory in flat space, to which they decay in the process of
tachyon condensation. This decay proceeds via an expanding bubble of flat
space embedded in a static sea of $C/Z_n$. Very recently, Vafa \vafa\ has
elegantly argued for the same picture by studying flows in gauged linear
sigma models and their mirror counterparts (see also
\refs{\DabholkarGZ,\DabholkarIF,\vafaatish,\chicago}).
 
The process of tachyon condensation seems even more mysterious in theories
that are not easily interpreted as finite excitations of  
supersymmetric backgrounds, as there are no clear candidates for the
endpoint of this process. Type 0A/B strings \refs{\seibwit,\DixonIZ} are
especially interesting tachyonic theories. These 10-dimensional
Lorentz invariant theories can be obtained from the supersymmetric type
IIA/B string theories by a $(-1)^F$ orbifold. Unlike
the (perhaps even more interesting) bosonic string, 
type 0 strings possess worldsheet supersymmetry and have 
R-R antisymmetric tensor fields and D-branes carrying R-R charges.  
The endpoint of tachyon condensation in these theories is of great
interest. Recently this question was given an intriguing twist 
in \costa\ where a duality was proposed between type 0A
string theory and a Kaluza-Klein-Melvin fluxbrane.
This duality suggests that type 0A theory is an excited state of IIA
theory (see \refs{\Suyamat,\Suyamath} for similar conjectures in heterotic
theories), and motivated the authors of \andya\ 
to conjecture that the endpoint of type 0A
tachyon condensation is the supersymmetric type IIA vacuum. 
Unfortunately, the duality of \costa\ is nonperturbative and so cannot be
easily used to verify the conjecture of of \andya. 

Rather than attempting a frontal assault on the problem of type 0 tachyon
condensation, in this paper we will try to outflank it by studying a large
class of type II backgrounds that includes the type 0 string theories as a
special limit. We will find it useful to study the so-called
``interpolating orbifold,'' in which type II strings are put on a circle of
radius $R$ with antiperiodic (Scherk-Schwarz) boundary conditions for the
spacetime fermions.\foot{A similar strategy was used to study the
non-supersymmetric heterotic strings in \refs{\BlumGW,\BlumCS}.}
In the limit $R\to\infty$, the boundary conditions
become unimportant and the theory returns to type II strings in flat
space. In the limit $R\to0$, on the other hand, the theory goes over to
type 0 strings in flat space (0B/A if we started with IIA/B) \atickwitten.
We will review this limit in subsection 2.1.

Even more room to maneuver can be had by introducing a second parameter
that interpolates between normal and Scherk-Schwarz boundary
conditions. Anti-periodicity for the fermions can be
imposed by accompanying the circle identification with a $2\pi$ rotation in
any orthogonal plane. An obvious generalization is therefore to choose an
orthogonal plane, and then to accompany the circle translation by a
rotation through an arbitrary angle $2\pi\zeta$. The resulting geometry is
flat ten-dimensional space $(\vec x\in R^{6+1},y\in R,z\in C)$, quotiented
by the isometry
\eqn\twisid{
(\vec x,y,z) \sim (\vec x, y+2\pi R,e^{2\pi i\zeta}z).
}
We will refer to such a space (or the non-trivial three-dimensional part of
it) as a ``twisted circle.'' Without loss of generality we assume
$0\leq\zeta\leq1$, with $\zeta=0$ corresponding to a normal circle and
$\zeta=1$ to a Scherk-Schwarz one.

The twisted circles form a moduli space of string backgrounds,
parametrized by $R$, $\zeta$, and the constant value $\Phi_0$ of the
dilaton. (Due to the supersymmetry breaking at non-zero $\zeta$, this is
only a moduli space at string tree level.) 
Tseytlin and Russo have extensively studied string propagation on these
backgrounds \refs{\tse,\TseytlinEI,\TseytlinZV,\RussoTF}.\foot{See also
\TakayanagiJJ. D-branes on twisted circles were discussed recently in
\refs{\DudasUX,\TakayanagiAJ}.}
In particular, they find that the
spectrum of the type II superstring includes a tachyon if and only if
\eqn\tachcond{
R^2 < 2\alpha'\zeta.
}
We review this spectrum in subsection 2.2.

According to \tachcond, there is a large range of parameters for which type
II theories on twisted circles are perturbatively unstable and decay via
tachyon condensation. In this paper we apply two methods to the problem of
finding the endpoint of this condensation.

In section 3, using the techniques of \refs{\vafa,\kentaro,\mirrorsym},
we construct a worldsheet conformal field theory that may be interpreted as 
a supersymmetric sigma model with a four real
dimensional target space.\foot{See \refs{\Suyamao,\piljin} for related work.} 
One of the directions in target space is effectively a Liouville direction.
At large positive values of the Liouville field, the three transverse
directions of the target space reduce to a twisted circle geometry with
a rational twist angle. At large negative values, on the other hand, they
reduce simply to $R^2\times S^1$. Assuming that motion in the Liouville
direction mimics the flow of the transverse three dimensional sigma model
under renormalization group flow, we conclude that under the tachyonic
perturbation it flows in the infrared
to the supersymmetric sigma model with target space $R^2\times S^1$.
Restated, we argue that the endpoint of 
tachyon condensation of a type II theory compactified on a twisted circle
(at least for rational values of the twist) is the same type II theory
compactified on an ordinary circle. 

A different approach to the question of tachyon condensation is to find a
zero-slope limit in which the tachyon is a mode of the supergravity
fields. We describe such a limit in section 4. The twisted circle in this
limit is most naturally described by T-dualizing, upon which it becomes an
H-flux 7-brane \refs{\dowkerb,\melvin,\gibbonsa, \RussoTF\BergmanRW}. The tachyon is a
Gregory-Laflamme-like instability of the fluxbrane, which we find
explicitly as a fluctuation of the NS-NS supergravity fields.
In this limit the problem of tachyon condensation is an interesting and
well posed problem in supergravity and
so may (for instance) be simulated on a computer. 
The study of tachyon condensation about H-flux branes, consequently, 
constitutes a check of the general arguments we present in section 3.
We set up this problem, and 
briefly report the inconclusive results of our rudimentary attempts at a 
numerical analysis of the relevant equations. 

In section 5 we end with a discussion of our results and their
implications.  In appendices A and B, we supply some technical details on the
tachyon and on RG flows. In appendices C and D we study the worldsheet
conformal field theory of strings on twisted circles. We determine the
tachyon vertex operator and compute its four point function. In appendix E
we use the results of appendix D to analytically study the initial dynamics 
of tachyon condensation when a fluxbrane is compactified on a circle
whose radius is chosen to make the tachyon almost massless.

\newsec{Review of type II strings on twisted circles}

\subsec{0B/A as IIA/B on a vanishing Scherk-Schwarz circle}

Consider type IIA strings on a Scherk-Schwarz circle of
radius $R$.
In the limit $R\to0$, all the states of non-zero momentum
around the circle are lifted out of the spectrum. This includes all the
fermions, since by the antiperiodic boundary conditions they have 
half-integral
momenta. In the zero-momentum sector, the effect of the boundary conditions
is to reverse the GSO projection for states of odd winding number $w$
(this follows from the modular invariance of the 1-loop partition
function \atickwitten).
Thus when
$w=2w'$ the spectrum consists of the
(NS+,NS+) and (R+,R-) sectors, and for $w=2w'+1$ the (NS-,NS-)
and (R-,R+) sectors. This is approximately the spectrum in the
zero-momentum, winding number $w'$ sector of 
type 0A theory on a circle of
radius $2R$. This equality becomes exact in the limit $R\to0$, where it is
more appropriate to describe the theory as 0B in infinite flat space.
It is similarly straightforward to show that
not only the spectrum but also the OPEs of the two theories
are the same in this limit. Hence we arrive at the statement that type
IIA compactified on a Scherk-Schwarz circle of zero size is the same
as type 0B at infinite radius.
By similar arguments, type IIB 
on a vanishing Scherk-Schwarz circle yields type 0A.

\subsec{Spectrum of type II strings on a twisted circle}

Type II strings propagating on twisted circles are described by a free
worldsheet theory, and so are easily quantized. The spectrum of the
resulting theory must be modified from that of type II strings on an
ordinary circle in several ways.

First, wave functions in the zero mode sector take the
form $e^{im\phi+i\alpha y}$, where $m$ is an integer as usual but
$\alpha = (k-m\zeta)/R$; $k$ is an integer (instead of the usual
$\alpha = k/R$). 

Next, note that the shortest winding number $w$ string that obeys $|z|>r$
has squared length $L^2 = (2\pi R)^2 + (2\pi r\zeta)^2$. Thus the world
sheet Hamiltonian includes a restoring harmonic oscillator potential in the
$z$-plane, which
confines all such states to the origin. In particular, the zero modes in
the $z$-plane of all winding modes are lifted.

The lifting of the zero modes in the $z$-plane is a special
case of a more general effect. In the winding number $w$ sector,
$z(\sigma,\tau)$ obeys the boundary conditions $z(2\pi,\tau) = 
e^{2\pi i\zeta}z(0, \tau)$, and so may be mode expanded as
\eqn\modexp{
z(\s,\tau) =
i\sqrt{\apm}e^{i\zeta y_0}\left(
\sum_{n+\zeta w\in Z}\beta_ne^{-in(\s+\tau)}+
\sum_{n-\zeta w\in Z}\tilde\beta_ne^{in(\s-\tau)}
\right).}
The shift in the moding of the $z$ oscillators results in a shift in their
level number, which it is not difficult to check is proportional to the
angular momentum: 
\eqn\nshift{
N_{z,\rm L}(\zeta) = N_{z,\rm L}(0) + \zeta wJ_{z,\rm L},
\qquad
N_{z,\rm R}(\zeta) = N_{z,\rm R}(0) - \zeta wJ_{z,\rm R},
}
where $J_{z,\rm L}$ and $J_{z,\rm R}$ are the contribution of left and
right movers, respectively, to the angular momentum in the $z$ plane
(recall that the angular momentum is the sum of occupation numbers of
`anticlockwise moving' modes minus the sum of occupation numbers of
`clockwise moving' modes). The change of moding in \modexp\ also results in
a modification of the zero-point energy in the NS sector.

Putting together these observations, we arrive at the following mass
spectrum \refs{\tse,\TseytlinEI,\TseytlinZV} in the NS-NS sector:
\eqn\spectrum{
M^2 = {2\over\apm}N_0 + \left({wR\over\apm}\right)^2
+ \left({m-\zeta J_{z,\rm L}-\zeta J_{z,\rm R}\over R}\right)^2 - {2\zeta
w\over\apm}(J_{z,\rm 
R}-J_{z,\rm L}-1).
}
Here $M$ is the mass relative to the remaining $6+1$-dimensional Lorentz
symmetry, while $N_0$ is the total level number at $\zeta=0$ (including
$-1/2$ for each NS sector, to make it an integer).
We note that \spectrum\ is only correct for $0 \leq \zeta \leq 1$. 
For instance, when $ 1 \leq \zeta \leq 2$, the formulae above remain valid if 
$\zeta$ is replaced by $\zeta'=2 - \zeta$. We will use this fact in section
3 below.

It is instructive to consider the special case $\zeta=1$. Since 
this involves compactification on a circle with a $2 \pi$ twist, one
might be tempted to conclude that the bosonic spectrum of this theory
is  identical to the spectrum of type II theory compactified on 
an ordinary circle. However, the shift in the moding of the $z$ oscillators
shifts the worldsheet fermion numbers by $\pm w$ as $\zeta$ is
taken from 0 to 1. This effectively
reverses the GSO projection in odd winding sectors, as was asserted in the
previous subsection.

It is evident from \spectrum\ that tachyons can only occur in the
winding sectors, and (for $w>0$) only if $J_{z,\rm R}-J_{z,\rm
L}>0$. Indeed, already at the lowest level allowed by the type II GSO
projections ($N_0=0$), there is an (NS+,NS+) state with $J_{z,\rm R}=1$,
$J_{z,\rm L}=-1$. In fact, this state, with $m=0$ and $w=1$, hence
$M^2=(R/\alpha')^2-2\zeta w/\alpha'$, is
the first state to become tachyonic as we increase $\zeta$. Consequently, 
a necessary and sufficient condition for the twisted circle to be tachyonic is
$2\apm\zeta>R^2$. As we take $\zeta$ to 1, this state flows to an
(NS-,NS-) state, and in the limit $R\to0$ it becomes the zero-momentum bulk
tachyon of the T-dual type 0 theory.

\newsec{Tachyon condensation from a linear sigma model analysis}

As  reviewed above, type II theories on twisted circles are tachyonic
if $R^2<2\alpha' \zeta$. The addition of a zero-momentum
tachyon to the world sheet conformal field theory induces a renormalization
group flow of that theory. In this section we will study this flow,
and argue that its endpoint is supersymmetric type II theory compactified
on an ordinary circle. 

Our arguments will be rather indirect. In
section 3.1 we will use an ${\cal N}=(2,2)$ linear sigma model to construct a
non-linear sigma model with a four (real) dimensional target space.
One of the target dimensions is a Liouville direction;
at large positive values of the Liouville field, the other three
dimensions reduce to the free conformal field theory describing the twisted
circle background, while at large negative values they reduce to the free
theory on $C\times S^1$.
We interpret motion in the Liouville direction as evolution under
renormalization group flow of the three dimensional sigma model when
perturbed by its tachyonic
operator, thereby concluding that it flows in the IR to
the flat $C\times S^1$ sigma model, restoring
space-time supersymmetry.

\subsec{The linear sigma model}

Consider an ${\cal N}=(2,2)$ U(1) gauged linear sigma model
 in two dimensions
\WittenYC\ containing, 
in addition to the gauge multiplet $\Sigma$, 
two charged chiral superfields
$\Ph_1$ and $\Ph_{-n}$ and an axionic chiral superfield $P$ with the
identification $P\sim P+2\pi i$.
Under U(1) gauge transformations the fields transform as
\eqn\ft{
\Ph_1 \to e^{i\a} \Ph_1,\qquad
\Ph_{-n} \r e^{-in \a} \Ph_{-n}, \qquad
P \r P + i \a .
}
The action for this system\foot{This system was analyzed in detail in
a beautifully written paper \kentaro. We follow the notations of that paper
in what follows.} is 
\eqn\act{S={1\over 2 \pi}
\int d^2\s d^4\t \left[{ \bar \Ph_1} e^V \Ph_1 
+{ \bar \Ph_{-n}} e^{-nV} \Ph_{-n}
+ {k \over 4} (P + {\bar P} + V)^2
-{1 \over 2 e^2} |\Sigma|^2 \right].}
For future reference we explicitly write the relevant parts of this action
in component fields using the Wess-Zumino gauge. The pieces of the action
involving the auxiliary gauge field $D$ are
\eqn\dterms{{1 \over 2 \pi} \int d^2\s 
\left( {D^2 \over 2 e^2}+D(|\ph_1|^2-n|\ph_{-n}|^2+k p_1) \right),}
where $\ph_1$, $\ph_{-n}$ and $p=p_1+i p_2$ refer to the lowest components of 
$\Ph_1$, $\Ph_{-n}$ and $P$ respectively. 
The kinetic terms for the scalar fields are
\eqn\kintsf{
{1 \over 2 \pi} 
\int d^2\s ( -\CD^\m \ph_1 \CD_\m \ph_1 -\CD^\m \ph_{-n} \CD_\m \ph_{-n}
- {k \over 2} \CD^\m p \CD_\m p),
}
where $\CD_\m \ph_1$, $\CD_\m \ph_{-n}$ are the usual covariant
derivatives, while $\CD_\m p=\p_\m p +i v_\m$ ($v_\m$ is the gauge
boson). It follows from the identification of $P$ and the normalization of the
kinetic term \kintsf\ that $p_2$ parameterizes a circle of radius
$R=\sqrt{\apm k}$.

In \act, \dterms, and \kintsf\ we presented the classical 
action for our system. Quantum mechanically this action is renormalized. 
Supersymmetry ensures that the renormalization of the D-term is 
independent of $e$ and so is a purely one loop effect \WittenYC.
The divergent contribution of one loop graphs to the $D$ term is 
\eqn\dterms{
{1 \over 2 \pi} \int d^2\s 
\left( D(1-n) \ln \Lambda \right).
}
This divergent contribution may be absorbed into a renormalization of 
$p_1$:
\eqn\renrep{
p_1 (\L)= p_1 (\mu)+{n-1\over k} \ln ({\L \over \mu}).
}
Notice that $p_1(\L)$ tends to $\infty$ in the UV and to $-\infty$ in the
IR.
 
We wish to study the dynamics of this model at low energies. The gauge 
multiplet of the theory is always massive (with  $m \geq e \sqrt{k}$) 
and may be integrated out. At sufficiently low energies we may restrict
attention to fluctuations on the supersymmetric 
manifold of zero-energy configurations satisfying
\eqn\ze{
|\ph_1|^2-n|\ph_{-n}|^2+k p_1=0,
}
modulo gauge transformations, so that the dynamics is captured by 
a two complex dimensional supersymmetric sigma model. In order to
explicitly study this sigma model, we will use the 
gauge freedom and \ze\ to solve for one of the three fields $(\ph_1,
\ph_{-n}, p)$. We will then plug this solution into the 
kinetic terms \kintsf\ and integrate out the gauge boson $v_\m$. 
This procedure is classical; however as we will see below, in the limit
$|p_1| \to \infty$ the gauge boson is infinitely massive. Since two
dimensional gauge theories are free in the UV, when $|p_1|$ is sufficiently
large the classical approximation is valid; we will carry out the analysis
only in this limit.

When $p_1 > 0$, $\ph_{-n}$ cannot be zero (see \ze), so we fix the gauge
by setting it real and positive. 
However, this gauge choice leaves unfixed a residual $Z_n$ group of
gauge transformations, generated by
\eqn\unfixexgt{
\ph_1 \r e^{{2 \pi i/ n}} \ph_1,\qquad
p_2 \r p_2 + {2 \pi \over n}.
}
\ze\ may then be used to solve for $\ph_{-n}$:
\eqn\phsolv{
\ph_{-n} = \sqrt{|\ph_1|^2 +k p_1 \over n}
= \sqrt{ k p_1 \over n}+ \CO\left(1\over\sqrt{p_1}\right).
}
As the ``Higgs'' field $\ph_{-n}$ has a vev of order $\sqrt{p_1}$, the
$v_\m$ gauge field is very massive for large $p_1$ and may now be
classically integrated out of \kintsf. In this limit
$v_\m = \CO(1/|\ph_{-n}|^2) = \CO(1/p_1)$ and so may be set to zero at leading
order. The kinetic term for $|\ph_{1}|$ is also suppressed compared to that
for $p_1$ by $\CO(1/p_1)$. Hence the dynamics in this limit is governed by
the flat sigma model
\eqn\ftds{S={1\over 2 \pi} \int d^2\s \left( 
\p_\m \ph_{1} \p^\m { \bar \ph_1}+{k \over 2} \p_\m p \p^\m { \bar p},
\right)
.}
where $\ph_1$ and $p$ are identified under the $Z_n$ group \unfixexgt.
Consequently, $\ph_1$ and $p_2$ parametrize a twisted 
circle of radius $R=\sqrt{\alpha'k}/n$ and twist angle
(in type II theories) $\zeta = 1+1/n$. (We will explain 
below why $\zeta=1+1/n$ rather than $1/n$.)

When $p_1$ is large and negative, the analysis 
presented above may be carried through with only minor modifications. 
A non-singular choice of gauge, in this case, is to set 
$\ph_1$ to be real and positive, which leaves no residual gauge
group. Solving for $\ph_1$ instead of $\ph_{-n}$, repeating the analysis
above yields the flat space sigma model
\eqn\ftdsn{S={1\over 2 \pi} \int d^2\s \left( 
\p_\m \ph_{-n} \p^\m { \bar \ph_{-n}}+{k \over 2} \p_\m p \p^\m { \bar p} 
\right).}
where $\ph_{-n}$ is an unconstrained chiral field and
$p_2$ is periodic with period $2 \pi$. Consequently
\ftdsn\ describes a sigma model on $C\times S^1\times R$ where the 
radius of the circle is $\sqrt{\apm k}$.

We now come to an important subtlety.  The central charge of this
low energy conformal field theory is not equal to 6, even though
it reduces in some regions to a sigma model on a flat four dimensional
target space. Recall that, under a change of scale, $p_1$ is additively renormalized 
(see \renrep). This implies that, for large $|p_1|$, the conformal field 
theory of the canonically normalized free boson 
$X\equiv{p_1 \sqrt{k \apm}}$ is governed by the linear dilaton system 
with charge $V=(1-n)/\sqrt{k \apm}$, and  central charge
$c=1+6 \apm V^2 = 1+6(n-1)^2/k$ (see equation 2.5.2 of \PolchinskiRQ\ for
notation). Consequently, the net central charge for the CFT is
$c=6+6(n-1)^2/k$. This is in precise agreement with the analysis given in
\kentaro.

In summary, the linear sigma model flows at low energies to a
$c=6+6(n-1)^2/k$ conformal field theory \kentaro. For
$ p_1 \gg 0$, this CFT reduces to the sigma model Liouville$\times$(twisted
circle). For $p_1 \ll 0$, it reduces to the sigma model 
Liouville$\times$($C\times S^1$).

\subsec{Mirror description}

We will now study the mirror \mirrorsym\ 
of the gauged linear sigma model defined in the previous section \kentaro.  
The mirror model is a Landau-Ginzburg model with two  
twisted chiral superfields $U$ and $Y_P$, and superpotential 
(see \refs{\kentaro, \vafa})
\eqn\twistesp{
\tilde W= e^{-nU}+ e^{-U-Y_P/n}.
} 
In the low energy limit, this model is conformal, and so both of the 
operators in the superpotential are marginal.  

According to the mirror map, ${\rm Re}\,Y_P \sim k {\rm Re}\,P$.\foot{More
precisely ${\rm Re}Y_P \sim k ({\rm Re}P+V)$. However, for $|p_1| \gg 1$
the gauge multiplet is very massive, and may be set to zero.} 
Consequently, in the asymptotic region $p_1\gg0$ the second 
term in the superpotential is negligible, and the Landau-Ginzburg model
reduces to the twisted circle times Liouville model.
The second operator is a marginal deformation of this theory, whose effect
vanishes for large $\Re Y_P$. It can be decomposed
into three factors: a twist operator $e^{-U}$ in the plane; a
winding operator $e^{-i{\rm Im}\,Y_P/n}$ on the circle; and a dressing
by $e^{-\Re Y_P/n}$. Together the first two factors
constitute a tachyon vertex operator on the twisted circle \twisid;
the third is a dressing by the Liouville mode,\foot{The dynamics of $Y_P$ is
governed by a Liouville theory \kentaro. In the rest of this section we
will often refer to $\Re Y_P$ as a Liouville direction} making it both
marginal and ${\cal N}=2$ supersymmetric.

Before ending this subsection we address two subtleties that we have 
ignored so far. Firstly, the worldvolume theory of a type 
II string is governed by a conformal field theory, modded out by a chiral 
GSO projection. As discussed in 
\refs{\aps\vafa\chicago}, it is possible to impose such a projection 
on the theory constructed above only if $n$ is odd.\foot{\refs{\aps\vafa
\chicago} actually discuss this for $C/Z_n$ but the arguments carry over 
to the CFT of this section.} Further, this projection removes 
fields with even twists, retaining all odd twisted fields, in particular 
the twist one tachyon of \twistesp. Secondly, the $Z_n$ identification 
\unfixexgt\ could be interpreted as acting on $\ph_{-n}$
as a rotation of either $2 \pi/n \times n$ or 
$2\pi(1+1/n)\times n$. Only the second of these actions 
acts trivially on spacetime fermions and is consistent with the chiral 
GSO projection. Consequently, in type II theory, we are forced to interpret 
the $\Re P \to\infty$ region of the CFT above as a twisted space with 
twist $\zeta=1+{1\over n}$ as we did in subsection 3.1 .

\subsec{Renormalization group flow}

In this subsection we will employ the CFT of subsection 3.1 in an argument
to predict the endpoint of tachyon condensation in type II theories on
twisted circles with twist $\zeta=1+1/n$ (equivalently $\zeta'=1-1/n$---see
under \spectrum).

Consider the twisted circle CFT whose central charge is $\hat{c}=3$,   
together with an independent CFT of central charge
$\hat{c}=6-{4(n-1)^2\over k}$. 
(The role of this theory is merely to provide additional central charge
to the system. It remains  a spectator through the action described
below.) Perturb the twisted circle CFT by the infinitesmal addition of a
tachyon, and formally couple this full system to two dimensional
supergravity. The Weyl mode of the two
dimensional graviton is dynamical, and combines with the original system to
form a conformal field theory with $\hat{c}=10$. 

We now study this $\hat{c}=10$ conformal field theory. Firstly note that
the Weyl field in this theory does not decouple from the perturbed  twisted
circle
CFT, as the tachyonic perturbation in the latter receives a Weyl dressing
that makes it marginal. However, as the Weyl field $\w$ represents the
scale factor of the twisted circle CFT, the effect of this
perturbation vanishes for $\w \to \infty$ (we choose
notation such that $\w \to \infty$ is the UV).  In this region the theory
reduces to the sum of a linear dilaton system for $\w$
($\hat{c}=1+{4(n-1)^2\over k})$) and the original conformal field theories,
together with an increasingly unimportant dressed tachyonic perturbation.
The system will be complicated for finite values of $\w$. Its behaviour 
at large negative values of $\w$ determines the endpoint of the RG flow 
initiated by a tachyonic perturbation in the twisted circle CFT. 

We now observe that the CFT of section 3.1 has (together with the spectator
CFT) all the properties expected of the CFT described in
the previous paragraph. The central charges of the two theories match by
construction.  $\Re Y_P$ has the correct conformal transformation
properties to be identified with $\w$. And at large values of $\Re Y_P$ it
reduces to the twisted circle CFT times a linear dilaton system, with a
dressed tachyonic perturbation. We thus propose that the endpoint of the
renormalization group flow initiated by the tachyon on the twisted circle
CFT of radius $R$ and twist $\zeta=1 +{1\over n}$
is determined by the $\Re  Y_P \to -\infty$ region of the CFT of subsection
3.1 and so is given by a CFT on a supersymmetric circle of radius $nR$. 

Given the formal nature of the above considerations, the following
observation might be useful. Suppose we are given an exact time dependent 
solution to the equations of motion of type II string theory starting at
$t=-\infty$ at the twisted circle theory, and evolving in time to the
endpoint of tachyon condensation. Analytically continuing this solution in
time (presuming that is allowed), yields a $\hat{c}=4$ conformal field
theory that interpolates between the twisted circle CFT and the endpoint
of tachyon condensation. The CFT of section 3.1 is such a theory  except that
its central charge $\hat{c}$ is greater than 4. The additional central
charge reflects the fact that the motion in the $\Re Y_P$ is flow in scale
rather than time.\foot{Indeed the linear dilaton provides a `dissipation'
term, allowing the motion to progress smoothly to the endpoint of tachyon
condensation. Evolution in time would conserve energy, and so would
presumably be considerably more complicated. The relation between Liouville
and time evolution was studied in
\refs{\DasDS,\PolchinskiFN,\DasDA,\BanksQE,\SchmidhuberBV}.}

We now return to an important detail of this interpretation. As we have
seen, the CFT of the previous sections for large $\Re Y_P$ is the 
product of the tachyon vertex operator and a Liouville dressing that keeps
the full operator marginal. Hence the dimension of the Liouville dressing
is $\zeta'-R^2/(2\apm)$.\foot{ This is easily directly verified in the UV
Liouville CFT. As we have not presented the action for $Y_P$ we work in
linear sigma model variables; recall that $\Re Y_P=  k \Re P$ for large
$\Re P$ according to the mirror map. The holomorphic scaling dimension of 
$$e^{-k p_1/n}= e^{\sqrt{k}X/\sqrt{\apm}n} $$
($X$ is the canonically normalized field defined in subsection 3.1)
is $$-{\apm \over 2}V{\sqrt{k} \over \sqrt{\apm} n}
-{\apm\over4}{k\over \apm n^2} = {n-1\over 2n} - {k \over 4 n^2}$$ 
where $V$ is the slope of the linear dilaton (see subsection 4.1.). 
Rewriting this expression in terms of the deficit angle and radius we 
find the scaling dimension $-R^2/4 \apm + \zeta'/2$.}
If $R^2>2\apm\zeta'$, so that the ``tachyon'' is not actually tachyonic,
then RG flow drives the surface on which the theory lives to larger values
of $\Re Y_P$, i.e. back to twisted circle. However, if $R^2<2\apm\zeta'$,
then RG flow drives the theory to smaller values of $\Re Y_P$. As discussed
in section 3.1 this means that at the endpoint the twist has
disappeared and we have a larger ordinary circle.

\subsec{Conclusions from this construction}

In this section we have argued that the endpoint of tachyon condensation 
for type II theory on a twisted circle with twist $\zeta=1+1/n$ 
and radius $R$ is the same theory compactified on an ordinary 
circle of radius $nR$. This result immediately leads to a prediction for 
the endpoint of tachyon condensation in type 0A/B theories. Recall that, 
for instance, uncompactified 
type 0B theory is type IIA theory on a twisted circle with 
$\zeta=1$ and $R \to 0$. In order to set $\zeta=1$ we take $n$ to infinity
holding $R$ fixed; the endpoint of tachyon condensation is IIA theory in 
uncompactified 10 dimensional space, for all values of $R$ including 
$R \to 0$. Consequently, it would appear that the endpoint of 0B/A tachyon
condensation is IIA/B theory in flat space!

Of course, taking this
limit could involve subtleties, since the localized tachyon of the twisted
circle becomes the delocalized tachyon of type 0 theory. Zamolodchikov
$c$-theorem-type arguments suggest that an RG flow seeded by such a tachyon
cannot end at a theory with the same central charge. Note, however, that we
are not strictly speaking studying RG flow here, but rather ``Liouville
flow.'' This type of evolution may more closely mimic dynamical evolution,
particularly insofar as it leaves the central charge invariant.

A generalization of the arguments of this section would imply that 
the final endpoint of tachyon condensation in IIA/B theory on a twisted 
circle of radius $R$ and with twist $\zeta=1+m/n$ is IIA/B theory 
on an ordinary circle of radius $nR$. Further, as any irrational number 
can be increasingly accurately approximated by rational numbers with 
increasing denominators, the endpoint of tachyon condensation of 
IIA/B theory on a twisted circle of irrational $\zeta$ would appear 
to be uncompactified IIA/B theory. Heuristically, tachyon condensation
unwinds the twist; this is achieved by a finite unwinding for rational 
$\zeta$ but an infinite unwinding for irrational $\zeta$.

\subsec{Supporting results}

The decay of $(C\times C)/ Z_{nm}$ orbifolds was studied in \refs{\aps,
\vafa, \chicago}. Let $Z_{nm}$ act by a rotation of $e^{2\pi i/m}$
on the first $C$ factor, and by a rotation of $e^{2\pi i/mn}$
on the second factor. In the limit $n\to\infty$, the second plane 
reduces to a cone with infinitesimal opening angle, and so (apart from the 
singularity at the origin) may be thought of as a zero size cylinder (this 
was noted in \aps\ in a very similar context). As the identification of
this orbifold links rotation around this cylinder to a twist about the
first $C$, this space approximates twisted cylinder of zero radius, with 
twist $1/m$. According to \vafa, the endpoint of tachyon
condensation in this background is $C\times (C/ Z_{n})$ where the $Z_n$ 
factor acts by rotations $e^{2\pi i/n}$, i.e. approximately 
an ordinary cylinder (still of zero radius) times an untwisted complex 
plane. This is in accordance with our our expectations described above.

\subsec{Open questions}

{\it RG flow versus time evolution}

Following several recent studies \refs{\aps, \vafa, \chicago, \vafaatish}, 
in this paper we have addressed tachyon condensation by studying
renormalization group flows on the world sheet of the string instead of the
actual on-shell process of tachyon condensation. While experience has often
shown (for instance in studies of open string tachyon condensation) that
these two processes have the same endpoint, it would be very interesting to
have a better understanding of the relationship between them. 

This question is especially delicate when the endpoint of RG flow is a CFT
with marginal deformations, as other solutions exist in the neighbourhood
of any such configuration. We have shown that RG flow (or more precisely
Liouville flow) starting from the twisted circle theory ends at flat space
on the smallest circle obtained after untwisting the original circle. The
size of this final circle (a modulus of the endpoint) is finite for
rational values of the twist, but infinite when the  twist is a generic
real number. It is noteworthy that the final circle depends so crucially on
the rationality or otherwise of the twist. It would be interesting to
understand if this feature persists under time evolution.

{\it Connection with energy}

Tachyon condensation in non-gravitational systems is a process that lowers
energy (after all the  excess energy has escaped to infinity in the form of
radiation). It is not clear to us how the energies of two systems with
different asymptotic geometries should be compared, but it would certainly
be very interesting to understand closed string tachyon condensation as a
process of energy minimization.
See \refs{\vafaatish, \chicago} for recent ideas in this direction. 

{\it The flow of the dilaton}

We have studied the RG flow of the string sigma model ignoring the
dilaton. The dilaton coupling to the string worlsheet is proportional to 
$\apm \int d^2 \s \sqrt{g}\Ph(X) R$, where $g$ is the worldsheet metric, 
$\Ph(X)$ is the dilaton, and $R$ is the worldsheet curvature scalar.  
Consequently, the string sigma model at tree level is independent
of the dilaton. 

The dilaton does modify the energy momentum tensor of the
world sheet theory, and so the change in $X$ under scale transformations is
given by 
\eqn\tranx{\delta X^\mu = -\apm \nabla^\mu \Phi \delta \ln \Lambda +{\rm higher\; order}.}
However, this effect  may be absorbed into a $\Lambda$ dependent coordinate 
($X^\m$) redefinition (we will see this in detail, in the supergravity
approximation, in subsection 4, see above eq. 4.8). 
Consequently, the renormalization group flow of the conformal field theory
may be decoupled from the dilaton and studied independently, as we have done in this section. 
 
On the other hand the behavior of the dilaton under RG flow certainly
depends on (in fact is determined by) the flow of the other fields in
the conformal field theory (this is explicit in the supergravity
approximation, as we will see in subsection 4.4). It would be very
interesting to study the evolution of the dilaton induced by the RG flows
of this paper as well as those described in \refs{\aps,\vafa}, and in
particular to determine the constant value of the dilaton at the endpoint
of tachyon condensation.

\newsec{Fluxbranes and the supergravity limit}

\subsec{A zero-slope limit}

In this section we will find a zero-slope limit of the models with twisted
circles. This limit is constructed so that the tachyon has a finite mass
as $\apm \to 0$ (of course generic stringy oscillators decouple in this
limit).

Consider the mass formula \spectrum. In order that states with winding
number $w \neq 0$ remain in the spectrum, $R$ must be scaled to zero along 
with $\apm$. We set $R=\apm/R'$ and hold $R'$ fixed when taking 
$\apm$ to zero. In order that the last term in \spectrum\ remains finite, 
we must also scale $\zeta$ to zero; we set $\zeta=q\apm/R'$,
and hold $q$ fixed. All states that are retained in this limit then have
$m=0$ and $N_0=0$.  
The spectrum of modes that survive this limit (in the NS-NS sector) is
\eqn\massformoth{
M^2 =({w \over R'})^2 +2wq ( s_L -s_R +l_L+l_R +1) 
+q^2(s_L+s_R+l_L-l_R)^2,
}\foot{For example, setting 
$s_L=-1$, $s_R=1$, $l_R=l_L=0$, \massformoth\ reduces to  
\eqn\tachmass{M^2 = k^2 - 2kq.}
where $k=w/R'$ is the momentum of the mode in this T-dual picture.}
Here $l_L$ and $l_R$ are the non negative occupation numbers of the zero
mode $z$  oscillators, and $s_L$ and $s_R$ the spins in the $z$ plane of the
GSO projected NS-NS ground state $b^R_{-\half} b_{-\half}^L |0\rangle$.  
This spectrum includes a tachyon if $qR'\leq 2$
(the most tachyonic mode has $w=1$, $s_L=-1$, $s_R=1$). For example, setting 
$s_L=-1$, $s_R=1$, $l_R=l_L=0$, \massformoth\ reduces to  
\eqn\tachmass{
M^2 = k^2 - 2kq.
}
where $k=w/R'$ is the momentum of the mode after T-duality.

\subsec{The H-flux tube}

The twisted circle geometry may be T-dualized on circles of constant 
$\tilde z \equiv e^{-i\zeta y/R}z$ \twisid\ to obtain an NS-NS
fluxbrane. The limit described above, together with an appropriate
scaling of the dilaton, i.e. 
\eqn\nicelimit{
R\to0,\qquad \zeta\to0,\qquad e^{\Phi_0}\to0,
}
with
\eqn\fixed{
q\equiv{\zeta\over R},\qquad e^{\Phi_0'}\equiv {\sqrt{\apm}\over R}e^{\Phi_0}
}
held fixed has a particularly natural description in this T-dual picture.
It is a flux 6-brane smeared in the $y$ direction
\eqn\background{\eqalign{
ds^2 = d\vec x^2+{1\over1+q^2r^2}dy^2+dr^2+{r^2\over1+q^2r^2}d\phi^2, \cr
e^{2\Phi} = {1\over1+q^2r^2}e^{2\Phi_0'}, \qquad 
B = {qr^2\over1+q^2r^2}dy\wedge d\phi,
}}
where $y$ is now the T-dual of the old $y$ direction (note $y \equiv y+ 2
\pi R'$) and $\tilde z =re^{i\phi}$. (In the classification of \ChenNR,
this is the $a=-1$, $p=6$,
$k=1$ fluxbrane.) The $r$-$\phi$ plane has a cigar geometry with
asymptotic radius $1/q$, and the curvatures and $H$ field strength are
concentrated near $r=0$.\foot{The geometry \background\ is
obtained after a T-duality in the limit \nicelimit,
\fixed, without necessarily also taking $\apm \to 0$. If in addition 
$q^2 \apm$ is taken to zero (the limit of subsection 4.1), supergravity is
a good approximation.}

To reiterate, the twisted circle models in the zero slope description of
the previous subsection are reliably described in supergravity as NS-NS 
H-Flux branes. It is easy to verify that \background\ obeys the
supergravity equations of motion. Since $\apm$ has been taken to zero, 
the entire spectrum \massformoth, including the tachyon, must be reproduced 
in type II supergravity about \background. 

It is easy to explicitly determine the tachyon as a fluctuation of
supergravity fields. We present our derivation in Appendix A; roughly
the linear fluctuation of the fields corresponding to the tachyon may be determined
by T-dualizing the polarization of the tachyon vertex operator. We find
\eqn\fluct{\eqalign{
\delta ds^2 &=
e^{iky-kqr^2/2}\left(
\left(dr-{iqr\over 1+q^2r^2}dy\right)^2 +
\left({r\over 1+q^2r^2}d\phi\right)^2
\right), \cr
\delta B &= ie^{iky-kqr^2/2}\left(dr-{iqr\over 1+q^2r^2}dy\right)\wedge
{r\over 1+q^2r^2}d\phi, \cr
\delta\Phi &= {e^{iky-kqr^2/2}\over 2(1+q^2r^2)}.
}}
(\fluct\ must be combined with the fluctuation related by $y\to-y$ to
obtain a real fluctuation of the fields.)
That \fluct\ is indeed a normal mode oscillation of the background
\background\ with mass given by \tachmass\ can be checked 
using the supergravity equations of motion. Note that the tachyon is 
confined to the region near $r=0$ by its $y$ momentum.

The fact that tachyon condensation in twisted models in the limit of
subsection 4.1 may be studied in supergravity is interesting, 
as it provides a method to test  the conclusions of the previous 
section, in a special case, using  a (perhaps numerical) supergravity
analysis. From another viewpoint it suggests an unfortunately rather
singular solution to the
endpoint of tachyon condensation about IIA/B flux branes in supergravity. 

In the next subsection we will briefly address the decay of fluxbranes in 
supergravity, and use supergravity intuition to suggest
possible endpoints. We will then observe that the prediction of subsec
3.5. infact predicts a singular endpoint. In the subsequent
subsection we will present a crude numerical analysis of the relevant
supergravity flows. 

\subsec{Flux tube decay: generalities}

This tachyon \fluct\ about \background\  has much in common with the
Gregory-Laflamme 
instability of black strings \greglaf.\foot{Similar observations were made
in \RussoNA.} Like the black string, \background\
is translationally invariant in one direction (the $y$ direction). Like our
tachyon \fluct, the Gregory-Laflamme instability about a black string is a
mode that carries momentum in this translationally invariant direction and
so breaks this translational invariance. The spectrum of the
Gregory-Laflamme tachyon as a function of the momentum in this direction  
 is qualitatively very similar to \tachmass. In particular, (as is the case
with for \tachmass), the Gregory-Laflamme tachyon is positive for
sufficiently large momentum. Consequently both our fluxbrane and the black
string are stable when compactified on sufficiently small circles. 

We recall standard lore about the Gregory-Laflamme tachyon condensation. 
Until recently it had been presumed that the Gregory-Laflamme instability,  
triggered at a particular wavelength, would lead to an array of black holes,
which would then by further instabilities coalesce into ever larger black
holes. However it was recently argued in \HorowitzCZ\ that such a motion is
not possible (the argument invoked mild assumptions). 
The authors of \HorowitzCZ\ argue that the black string instead
settles down into a stable (and as yet unknown) ``string of pearls''
configuration, at a preferred wavelength set by its Schwarzschild
radius. 

The analogy between \background\ and the black string might lead one to
explore the corresponding two endpoints to the decay of \background. 
The first is the SO(3)-symmetric NS flux 6-brane of
\refs{\andya,\CostaIF,\SaffinKY}, which carries
an infinite amount of flux and has a conical geometry at infinity. The
second is as yet unknown periodic arrangement of flux in the
y-direction, which would not change the boundary conditions at infinity.

The analysis of the previous section, however, predicts a third and rather
singular endpoint to this evolution. 
Before T-dualizing, the fluxbrane in IIA/B is described by the limit
\nicelimit\ together with $\apm \to 0$, i.e. 
$R=\apm/R'$, $\zeta=q\apm/R'$ with $\apm \to 0$ and $q$,
$R'$ fixed. As the denominator of $1-\zeta$
for small $\zeta$ is greater than or equal to $1/\zeta$, according 
to section 3.5 the endpoint of tachyon condensation is IIA/B theory on a
circle of size at least $1/q$. T-duality relates this to IIB/A theory
on a circle of size smaller than $\apm q=0$ in the limit under consideration.
The results of section 3.5 thus predict that the supergravity solution
\background\ will evolves into  IIB/A on a zero size circle, i.e. a
singular geometry. It is of great interest to verify this prediction
directly in supergravity.

\subsec{Flux tube decay: numerics}

In our primitive attempts at a numerical simulation of the supergravity
equations, we did not try to model the dynamical time evolution, but
instead attempted to simulate worldsheet RG flow. 
 The RG equations on the NS-NS fields are:
\eqn\rgflowgen{\eqalign{ 
{ { \dot G}_{\m \n} \over \apm} & =-R_{\m\n}- 2 \nabla_\m \p_\n 
\Phi +{ 1\over 4} H_{\m \a \b}{H_\n}^{\a\b} + \nabla_\m\xi_\n
+\nabla_\n\xi_\m, \cr
{ { \dot B}_{\m \n} \over \apm} & =  \half \nabla^\alpha H_{\alpha\mu\nu}
-H_{\alpha\mu\nu}\p^\alpha\Phi+ H_{\m\n\rh}\xi^\rh, \cr
{ {\dot \Phi} \over \apm} &=\half \nabla^2 \Phi- \p_\mu\Phi\p^\mu\Phi
+{1 \over 4}
|H|^2 + \partial_\m\Phi \xi^\m.}}
The dots on the left hand side 
 denote derivatives with respect to the logarithm of the
renormalization scale. The vector $\xi$ performs small spacetime
diffeomorphisms along with RG flow \aps, allowing, for example, the
imposition of  a gauge-fixing condition on the flow.  
Choosing $\xi_\mu \r \partial_\mu\Phi+\xi_\m$
eliminates the dilaton \aps\ from the first two equations in \rgflowgen. In
order to determine the flow of the dilaton, however,  one needs all the
other fields (this illustrates the general principle explained in
subsection 3.7).  Since our problem is effectively three dimensional 
 is possible to express the degree of freedom in the antisymmetric
tensor $B_{\mu\nu}$  in terms of a scalar by defining
$H_{ry\phi}=\epsilon_{ry\phi}\psi$. 
The RG flow equations reduce to 
\eqn\rgflowpsi{\eqalign{ 
{ { \dot G}_{\m \n} } &=  -R_{\m\n} +{ 1\over 2} G_{\m\n} \psi^2
+\nabla_\m\xi_\n 
+\nabla_\n\xi_\m, \cr
{ { \dot \psi } } &=   \half \nabla^2 \psi 
+ \half \psi R -{3\over 4} \psi^3+\partial_\m\psi \xi^\m, \cr 
{ {\dot \Phi} } &=\half \nabla^2 \Phi+{1 \over 4}
\psi^2 +\partial_\m\Phi \xi^\m.
}}

In order to study these equations we imposed a gauge in which the metric
was diagonal 
with $g_{\phi\phi}$ fixed, which, given the symmetries of the problem,
leaves no remaining coordinate freedom.  Our simulations initially
reproduced the  linear evolution of the normal mode, with the predicted
eigenvalue, but upon entering the nonlinear regime of evolution quickly
ran into a singularity. However, we are unsure whether the singularity that
we saw in our numerics was that described the previous subsection, or
simply than a prosaic coordinate singularity due to an unlucky choice of
gauge. We present some additional details of these numerical simulations in
Appendix B.

\newsec{Discussion}

In this paper we have studied the compactification of type II string
theories on twisted circles and the condensation of tachyons in this
system. We have argued that the final endpoint of tachyon condensation in 
these models is type II theory on an untwisted circle. In particular, if
our analysis can be trusted in the limit that the system becomes type 0A(B)
theory, then it would predict that the endpoint of tachyon condensation in
those theories is type IIB(A) theory in flat space.

While we believe that the arguments of this paper have made a strong case
for this conclusion, our arguments have been indirect and several questions
remain open. For instance we have not been able to follow the flow of the
dilaton, in the passage from twisted to ordinary circles, and so cannot
predict its final value. It would certainly be useful to understand the
tachyon condensation process in detail. It is thus especially interesting
that in the zero-slope limit described in subsection 4.1 behind, tachyon
condensation may be analysed in supergravity. A detailed (perhaps
numerical) study of the supergravity evolution might prove very
illuminating.

The results of this paper fall into the simple
pattern observed in other recent studies of tachyon condensation. In the
situations studied so far, it appears that closed string tachyons
generically represents the  instability of an excited state to decay into a 
supersymmetric ground state of M theory. It would
be fascinating to learn if this more generally true in string 
theories with tachyons and in particular where the bosonic string fits into
this story.

An intriguing feature of our analysis is the appearance of an auxiliary 
spacetime dimension that could be interpreted as a scale. This 
is reminiscent of the holographic nature of the AdS/CFT duality, and 
in particular the duality of little string theories to linear dilaton 
string theories. In fact, the auxiliary CFT constructed in section 3 of
this paper is closely related to little string theories compactified
on a twisted circle.\foot{ We would like to thank J. Maldacena for bringing
this to our attention.} It would be interesting to understand these
connections better. It would also be very interesting to explore the
applications of our results to the Hagedorn transition in finite
temperature string theory.

Apart from the perturbative instabilities, type II theories on twisted
circles also have nonperturbative instabilities which correspond to the
nucleation of branes in a strong background field. This
is reminiscent of the situation in brane-antibrane systems where for
large brane separation the open string tachyons disappear
\refs{\BanksCH,\GreenUM} and are replaced by a nonperturbative bounce eating
up the branes \refs{\callan,\horava}.
In \andya\ it was conjectured that the ten dimensional R-R Melvin flux
7-brane can decay to supersymmetric type IIA. In this paper we
analyzed related systems where the flux is in the NS-NS sector and can
therefore be analyzed in string perturbation theory. Even in this case
there are nonperturbative instabilities, and the results of
\EmparanGM\ seem to indicate that perturbative tachyon
condensation and nucleation of KK-branes have the same endpoint.

\vskip 0.4in

\centerline{\bf Acknowledgements}

\vskip .1in

We would like to thank K. Hori and C. Vafa for explaining their work, and
for several extremely useful discussions. We would also like to thank
R. Gopakumar for useful discussions and comments on the manuscript, and
M. Brenner and A. Mody for advice on numerical simulations. We also
benefited from discussions with
A. Adams, M. Aganagic, T. Banks, A. Dhabolkar, S. Giddings, J. Harvey,
G. Horowitz, A. Karch, J. Maldacena, A. Mikhailov, J. Polchinski,
M. Rangamani, N. Seiberg, A. Sen, E. Silverstein, A. Strominger,
T. Takayanagi, N. Toumbas and T. Uesugi.
JRD would like thank the high energy group at Harvard for hospitality where
part of this work was done. His work is supported in part by NSF grant
PHY00-98395. The work of MG was supported in part by  NSF-PHY/98-02709. The
work of MH and SM was supported in part by DOE grant DE-FG01-91ER40654. The
work of SM was also supported  in part by a Harvard Junior Fellowship.

\appendix{A}{The tachyon in supergravity}
In section 2 it was shown that tachyons for a string on twisted circles 
 have   winding number  $w= \pm k$ and  spin
$j_l=\pm 1,j_r=\mp  1$. The zero modes are
 replaced by a set 
of creation and annihilation operators of a two dimensional harmonic
oscillator. The tachyon will be in the ground state wave function of this
oscillator. Hence the linear fluctuations of the graviton 
and antisymmetric tensor
associated with the $w=\pm k$ tachyon is given by
\eqn\tachverta{h_{rr}= e^{-k q r^2/2},\quad h_{\phi\phi}=  
 r^2 e^{-k q r^2/2},
\quad b_{r\phi}= \pm i r e^{-k q r^2/2}.}

There is a subtlety in the identification of the polarization of the vertex
operator and the fluctuations of the (flat) supergravity
background. In 
string perturbation theory the graviton
and dilaton satisfy a generalized de Donder gauge condition
\eqn\dedonder{\partial_{\mu} h^\mu_\nu-{1\over 2} \partial_\nu h^\mu_\mu+2
\partial_\nu \Phi=0}
Since we do not have any momentum in the $r,\phi$ directions this condition
can be satisfied if $\Phi = h^\mu_\mu/4$; hence in addition to \tachverta\
the tachyon wave functions also contains a  dilaton fluctuation given by
\eqn\tachvertc{
\delta\Phi= {1\over 2} e^{-k q r^2/2}.
}

The physical tachyon is given by the real combination of the $w=+k$ and $w=-k$
wave functions of the tachyon. There are thus  two tachyon states,
because we are on a  circle, as $R\to 0$ (or
$R' \to \infty$ in the T-dual picture) the two states correspond to
positive or negative momentum.

The $r,\phi$ coordinates are flat but have nontrivial identifications, it
is therefore useful to express the the polarizations in the
single valued  (untwisted coordinates), using $\tilde \phi= \phi
 +q y$.

\eqn\melback{\eqalign{h_{yy}&= q^2 r^2 e^{-k q r^2/2},\quad \quad
h_{rr}= e^{-k q r^2/2},\cr
h_{y\tilde\phi}&=  qr^2 e^{-k q r^2/2},\quad \quad
h_{\tilde\phi\tilde\phi}=  r^2 e^{-k q r^2/2},\cr
b_{r\tilde\phi} &=i  r e^{-k q r^2/2},\quad \quad
b_{ry} =\pm i  q r e^{-k q r^2/2},\cr
\delta \Phi&= {1\over 2} e^{-k q r^2/2}
.}}
These are small fluctuations around the background of the twisted circle.
 We now want to
perform a T-duality along the y-circle. The winding mode $w=\pm k$
 is turned into a momentum mode. The background and the small fluctuations
can be obtained 
using the standard Buscher duality rules \BuscherQJ\  and expanding to order
linear order in the small fluctuations.  The linearized fluctuations
are given by 
\eqn\melbackb{\eqalign{ h_{yy}&= -
{q^2r^2\over ( 1+q^2 r^2)^2} e^{-k q r^2/2} e^{\pm ik y},\quad\quad
 h_{rr}=  e^{-k q r^2/2} e^{\pm ik y},\cr
 h_{yr}&= \mp i { q r\over1+q^2 r^2} e^{-k q r^2/2} e^{\pm ik y},\quad \quad 
\tilde h_{\tilde \phi\tilde \phi}=  {r^2 \over (1+q^2 r^2)^2} e^{-k q r^2/2}
e^{\pm ik y},\cr 
\tilde h_{y\tilde \phi}&= 0,\quad \quad
\tilde h_{\tilde \phi r}= 0,\quad \quad  b_{yr}= 0, \cr
\tilde b_{r\tilde \phi} &= \pm i {r\over 1+q^2 r^2} e^{-k q r^2/2} e^{\pm ik y},\quad \quad
\tilde b_{y\tilde \phi} = {q r^2\over (1+q^2
r^2)^2}  e^{-k q r^2/2} e^{\pm ik y},\cr
\delta\Phi &= {1\over 2}  {1 \over 1+q^2r^2} e^{-k q r^2/2} e^{\pm ik y}.
}}
It is easy to check that   \melbackb\ corresponds a normal mode  of the
linearized equations of motion growing like $e^{ m t}$  where $m$ is the
mass of the tachyon $ m = \sqrt{2kq-k^2}$, however   it would be a
formidable task to find \melbackb\ without the help
of T-duality.

The tachyon wave function is localized near the center of the fluxbrane in a
region $r< 1/|k q|$, note that this region becomes larger for smaller
tachyon momenta $k$.  It is interesting that the tachyon has nonzero
momentum in the $y$ direction, since one naively would expect that
adding  momentum raises the mass of a state.  The condensation of the
tachyon will therefore break the translation invariance along the $y$
circle of the background.

\appendix{B}{Details of the numerics}

We performed a numerical analysis of the supergravity RG flow equation
by fixing to a particular gauge. After eliminating the dilaton from 
the flow equations for the other fields (as described in Section 4).
We simplify the general flow equations
\eqn\rgflownn{\eqalign{
\dot g_{ij} &=
- R_{ij} + \partial_i\sigma\partial_j\sigma + \nabla_i\partial_j\sigma
+ \half g_{ij}\psi^2 + \nabla_i\xi'_j+\nabla_j\xi'_i, \cr
\dot\sigma &= 
\half\nabla^2\sigma + \half\partial_i\sigma\partial^i\sigma
+ {1\over4}\psi^2 + {\xi'}^i\partial_i\sigma, \cr
\dot\psi &=
\half\nabla^2\psi+\half\partial_i\psi\partial^i\sigma
+ \psi(\half R-\partial_i\sigma\partial^i\sigma-\nabla^2\sigma) -
{3\over4}\psi^3 + {\xi'}^i\partial_i\psi, \cr
\dot\Phi &=
\half\nabla^2\Phi + \half\partial_i\Phi\partial^i\sigma + {1\over4}\psi^2
+ {\xi'}^i\partial_i\Phi.
}}
by choosing $\xi$ so that the flow respects the gauge 
(from this point we set $q=1$ for simplicity):
\eqn\gaugedef{
ds^2 = e^{2\eta}dy^2+e^{2\alpha}dr^2+e^{2 \s} d \ph^2, \qquad
\sigma = \ln{r\over\sqrt{1+r^2}}.
}
In order for evolution to respect this gauge, $\xi$ must be chosen such
that $\dot\sigma=\dot g_{ry}=0$. We may
solve these equations for $\xi'$:
\eqn\xidef{\eqalign{
{\xi'}^r &= \half e^{-2\alpha}
\left(\partial_r\alpha-\partial_r\eta+{3r\over1+r^2}\right) -
{1\over4}r(1+r^2)\psi^2,
\cr
{\xi'}^y(r') &= -\int_{r'}^\infty dr\,e^{-2\eta}\left(
{\partial_y\alpha\over
r(1+r^2)}-e^{2\alpha}\partial_y{\xi'}^r
\right).
}}
(Note that the integral in ${\xi'}^y$ should be done from infinity in order
to leave the asymptotic region invariant .)
The RG flow equations in this gauge then become:
\eqn\rgflownnn{\eqalign{
\dot\eta &=
- {1\over4}R + {e^{-2\alpha}\partial_r\eta\over2r(1+r^2)} +
 {1\over4}\psi^2 + \partial_y{\xi'}^y + {\xi'}^y\partial_y\eta
 + {\xi'}^r\partial_r\eta, \cr
\dot\alpha &=
- {1\over4}R -
\half e^{-2\alpha}
\left({3\over(1+r^2)^2}+{\partial_r\alpha\over r(1+r^2)}\right) +
{1\over4}\psi^2 + \partial_r{\xi'}^r + {\xi'}^y\partial_y\alpha
 + {\xi'}^r\partial_r\alpha, \cr
\dot\psi &=
\half e^{-2\eta}
\left(\partial_y^2\psi+(\partial_y\alpha-\partial_y\eta)\partial_y\psi\right)
+\half e^{-2\alpha}\left(\partial_r^2\psi +
\left(\partial_r\eta-\partial_r\alpha+{1\over r(1+r^2)}\right)
\partial_r\psi \right) \cr
&\qquad\qquad\qquad
+ \psi\left(
\half R+e^{-2\alpha}\left({3\over(1+r^2)^2} +
{\partial_r\alpha-\partial_r\eta\over r(1+r^2)}\right) - {3\over4}\psi^2
\right)
+ {\xi'}^r\partial_r\psi + {\xi'}^y\partial_y\psi,
}}
where the Ricci scalar $R$ is given by
\eqn\ricci{
R = 
2e^{-2\eta}\left(\partial_y\alpha\partial_y\eta-(\partial_y\alpha)^2
-\partial_y^2\alpha\right) +
2e^{-2\alpha}\left(\partial_r\alpha\partial_r\eta-(\partial_r\eta)^2
-\partial_r^2\eta\right).
}
We numerically solved the gauged fixed equations \rgflownnn\ using an
explicit time stepping algorithm, solving for the gauge fixing constraints 
\xidef\ at each time step and imposing the appropriate boundary conditions
for $r=0$ and a large $r$. The initial conditions were giving by the
background \background\ plus a small tachyonic perturbation \melbackb.

\appendix{C}{Twisted circles on the worldsheet}

\subsec{CFT description}
We will review the mode expansions 
of twisted circles on the world sheet. We use 
Euclidean world sheet
conventions which makes it convenient to apply conformal field theory
methods. In the discussion below we will set $\alpha' =2$. 

The bosonic part of the world sheet action is given by
\eqn\action{
S= \frac{1}{4\pi} \int d^2 z \partial X^\mu \bar{\partial} X_\mu.
}
Let us restrict our attention  
to the coordinates of the 2-plane which have twisted
boundary conditions in  Type
IIA.
In terms of free fields 
$X^+ = X^1 + iX^2$ and $X^-  = X^1 -i X^2$ 
the twisted circle background  satsifies the following
boundary conditions
\eqn\bc{\eqalign{
X^+ (e^{2\pi i}z, e^{-2\pi i} \bar{z} ) &= e^{2\pi i \nu} X^{+}(z,
\bar{z} ), \cr
X^- (e^{2\pi i}z, e^{-2\pi i}\bar{z}) &= 
e^{-2\pi i \nu} X^{-} (z,
\bar{z}) 
.}}
Here $\nu = w\zeta = wq R$ with $0<\nu <1$.
The mode expansions for these fields are given by
\eqn\expansion{\eqalign{
X^+(z, \bar{z}) = 
i\sqrt{2} \sum_n \left( 
\frac{1}{n-\nu}\frac{\alpha_{n-\nu} ^{+}}{z^{n-\nu}} + 
\frac{1}{n+\nu} \frac{\tilde{\alpha} ^+_{n+\nu} }{\bar{z}^{n+\nu}}
\right),
\cr
X^-(z, \bar{z}) = 
i\sqrt{2} \sum_n \left( 
\frac{1}{n+\nu}\frac{\alpha_{n+\nu} ^{-}}{z^{n+\nu}} + 
\frac{1}{n-\nu} \frac{\tilde{\alpha} ^-_{n-\nu} }{\bar{z}^{n-\nu}}
\right)
.}}
The canonical commutation relations for the modes are given by
\eqn\commute{\eqalign{
[\alpha^+_{n-\nu}, \alpha^-_{m+\nu}] = (n-\nu) \delta(n+m), \cr
[\tilde{\alpha}^+_{n+\nu}, \tilde{\alpha}^-_{m-\nu}] = (n+\nu) 
\delta(n+m) 
.}}
Substituting the mode expansions \expansion\ in the definition of the
zero modes $L_0$ and $\bar{L}_0$ of the stress energy tensor and using
the commutations relations \commute\ we obtain
\eqn\stressm{\eqalign{
L_0 &= \sum_{n=1}^\infty (\alpha^-_{-n+\nu} \alpha^+_{n-\nu}+
\alpha^+_{-n-\nu} \alpha^{-}_{n+\nu}) + \alpha^+_{-\nu} \alpha^-_\nu
-\frac{1}{12} + \frac{1}{2}\nu(1-\nu), \cr
\bar{L}_0 &= 
\sum_{n=1}^\infty ( \tilde{\alpha}^+_{-n+\nu} \tilde{\alpha}^-_{n-\nu}
+ \tilde{\alpha}^-_{-n-\nu} \tilde{\alpha}^+_{n+\nu}) +
\tilde{\alpha}^-_{-\nu}\tilde{\alpha}^+_\nu
-\frac{1}{12} + \frac{1}{2} \nu(1-\nu) 
.}}
Here $L_0, \bar{L}_0$ is written with the fractional moding such that
the linear term which occurs with $j_L -j_R$ is not present. It can be
easily shown that this is the same as \nshift.

The operator which creates the twisted sector out of the vacuum is
given by $\sigma_\nu (z, \bar{z})$. It is defined  by the following
operator product expansions  \DixonQV. 
\eqn\ope{\eqalign{
\partial X^+ (z)\sigma_\nu(w, \bar{w}) = \frac{\tau_\nu (w,
\bar{w})}{(z-w)^{1-\nu}}, \;\;\;&\;\;\;
\partial X^- (z)\sigma_\nu(w, \bar{w}) = \frac{\tau_\nu^\prime (w,
\bar{w})}{(z-w)^\nu}, \cr
\bar{\partial} X^+ (\bar{z}) \sigma_\nu(w, \bar{w}) = 
\frac{\tilde{\tau}_\nu^\prime (w,
\bar{w})}{(\bar{z}-\bar{w})^\nu}, \;\;\;&\;\;\;
\bar{\partial} X^- (\bar{z}) \sigma_\nu(w, \bar{w}) = 
\frac{\tilde{\tau}_\nu (w,
\bar{w})}{(\bar{z}-\bar{w})^{1-\nu}} 
.}}
These equations also 
define the excited twist operators $\tau$'s. There are similar
operator product expansions 
for the anti-twist operator $\sigma_{-\nu}(z, \bar{z})$.
The conformal dimension of the twist operator 
$\sigma_\nu$ or the anti-twist operator $\sigma_{-\nu}$ 
is $h=\bar{h} = \frac{\nu(1-\nu)}{2}$. 
Note that this is the contribution of the zero point
energy due to the fractional moding in \stressm.

The world sheet action for the fermions is given by
\eqn\faction{
S =\frac{1}{4\pi} \int d^2 z \left( \bar{\partial} \psi^\mu \psi_\mu -
\tilde{\psi}^\mu \partial \tilde{\psi}_\mu \right). 
}
Using the suspersymmetric variation on the \expansion\ we find the
modes expansion of the superpartners of $X^+$ and $X^-$.
\eqn\fexpansion{\eqalign{
\psi^+(z) = i\sqrt{2} \sum_{r} \frac{\psi^+_{r-\nu}}{z^{r-\nu+1/2}},
\;\;\;&\;\;\;
\psi^-(z) = i\sqrt{2} \sum_{r} \frac{\psi^-_{r+\nu}}{z^{r+\nu+1/2}}, \cr
\tilde{\psi}^+(\bar{z}) = 
i\sqrt{2} \sum_{r} 
\frac{\tilde{\psi}^+_{r+\nu}}{\bar{z}^{r+\nu+1/2}}, 
\;\;\;&\;\;\;
\tilde{\psi}^-(\bar{z}) = 
i\sqrt{2} \sum_{r} 
\frac{\tilde{\psi}^-_{r-\nu}}{\bar{z}^{r-\nu+1/2}} 
.}}
Here $r$ is half integral in the Neveu-Schwarz sector and integral in
the Ramond sector. We will focus on the Neveu-Schwarz sector therefore 
for the rest of the discussion $r$ is half integral.
The anti-commutation relations are given by
\eqn\anticommute{\eqalign{
\{ \psi^+_{r-\nu}, \psi^-_{s+\nu} \} &= \delta(r+s), \cr
\{ \tilde{\psi}^+_{r+\nu}, \tilde{\psi}^-_{s-\nu} \} &= \delta(r+s) 
.}}
Substituting the expansion \fexpansion\ for the zero mode of the
stress energy tensor  and using
\anticommute\ we obtain
\eqn\fstressm{\eqalign{
L_0 &= \sum _{r>0} (r-\nu) \psi^-_{-r+\nu}\psi^+_{r-\nu} + 
\sum_{r>0} (r+\nu) \psi^+_{-r-\nu}\psi^-_{r+\nu} 
-\frac{1}{24} + \frac{\nu^2}{2}, \cr
\bar{L}_0 &= 
\sum_{r>0} (r-\nu) \tilde{\psi}^+_{-r+\nu} \tilde{\psi}^-_{r-\nu}
+ \sum_{r>0} (r+\nu) \tilde{\psi}^-_{-r-\nu} \tilde{\psi}^+_{r+\nu}
-\frac{1}{24} +\frac{\nu^2}{2}
.}}
Here to make our discussion less cumbersome we have assumed
$0<\nu<1/2$. The summation over $r$ in the above expressions run over
the positive half integers.

To construct the fermionic twist operators 
we first bosonize the fermions by
defining
\eqn\bosonize{\eqalign{
\psi^+ (z) = i\sqrt{2} e^{iH(z)}, \;\;\;\; \psi^- (z) = i\sqrt{2}
e^{-iH(z)}, \cr
\tilde{\psi}^+(\bar{z}) = 
i\sqrt{2} e^{i\tilde{H}(\bar{z}) }, \;\;\;\; \tilde{\psi}^-(\bar{z}) 
= i\sqrt{2} e^{-i\tilde{H}(\bar{z}) } 
.}}
The twist operator is given by $e^{i\nu (H(z)-\tilde{H}(\bar{z}) )}$.
Note that the dimension of this twist operator is the same as the
zero point energy of the fermions due to the fractional modding in
\fstressm. From \stressm\ and \fstressm\ it is easy to see that the
full zero point energy including the remaining $3$ complex untwisted
coordinates is given by $\frac{\nu}{2} -\frac{1}{2}$.

Due to the twisted circle  boundary conditions \twisid\ the zero
modes of the $Y$ coordinate are quantized as follows.
\eqn\yquant{\eqalign{
p_L + p_R &= 2(\frac{n}{R} - \frac{(j_L +j_R) \nu }{R}), \cr
p_L -p_R &= Rw
.}}
The $L_0$ and $\bar{L}_0$ contribution from these zero modes is
$\frac{p_L^2}{2}$ and $\frac{p_R^2}{2}$ respectively.

\subsec{The tachyon vertex operator}

We now construct the vertex operator corresponding to the tachyon
which survives in the zero slope limit.
Consider the following state over the twisted vacuum
\eqn\tachst{
\psi^-_{-1/2 +\nu} \tilde{\psi}^+_{-1/2 +\nu} |Twist, w \rangle.
}
Let us write this as a vertex operator in the $(-1, -1)$ picture,
\eqn\tachvertex{
\int d^2 z e^{-\phi(z)} e^{-\tilde{\phi}(\bar{z})}
\sigma_{\nu}(z, \bar{z}) e^{i (\nu-1)H(z)} e^{-i(\nu
-1)\tilde{H}(\bar{z}) }
e^{i\frac{wR}{2} (Y(z) - \tilde{Y}(\bar{z}) }
.}
Here $\phi(z)$ and $\tilde{\phi}(z)$ are the superconformal ghosts.
This state has $j_L + j_R=0$. 
Note that the conformal
dimension of this operator is given by
\eqn\dim{
L_0 = -\frac{\nu}{2} + \frac{w^2R^2}{8} \;\;\;\;
\bar{L}_0 = -\frac{\nu}{2} + \frac{w^2 R^2}{8} 
.}
This gives the following mass for the field
\eqn\mass{
M^2 = \frac{1}{4} (wR -2q)^2 - q^2
,}
where $\nu =w\zeta= wqR$. This agrees with the mass formula 
\tachmass. 
Note also there is a tachyon from the anti-twist operator
corresponding to the state in the opposite winding sector.
We write down its vertex operator,
\eqn\tachvertexa{
\int d^2 z e^{-\phi(z)} e^{-\tilde{\phi}(\bar{z})}
\sigma_{-\nu}(z, \bar{z}) e^{-i (\nu-1)H(z)} e^{+i(\nu
-1)\tilde{H}(\bar{z}) }
e^{-i\frac{wR}{2} (Y(z) - \tilde{Y}(\bar{z}) }
.}
This state also has the same mass given by \mass.

\appendix{D}{The tachyon potential near the critical radius}

In this appendix we evaluate 
the tachyon potential in the zero slope limit for a tachyon close to
marginality.
We use  strategy developed in \GavaJT\ for evaluating the tachyon
potential.
We first evaluate the 4-pt tachyon on shell
amplitude. 
By factorizing this amplitude along various channels we can extract the
cubic couplings involving two tachyons and other fields which
survives the zero slope limit. 
Subtracting the contribution of the massless exchanges in the 4-pt
tachyon amplitude we can extract the leading 
contact quartic interaction term
involving the tachyons.
\subsec{The 4-point tachyon on-shell amplitude}

The tachyon vertex operator in the $(-1,-1)$ picture is given by
\eqn\tvertex{
{\cal T} =  e^{-\phi(z)} e^{-\tilde{\phi}(\bar{z})}
\sigma_{\nu}(z, \bar{z}) e^{i (\nu-1)H(z)} e^{-i(\nu
-1)\tilde{H}(\bar{z}) }
e^{i\frac{wR}{2} (Y(z) - \tilde{Y}(\bar{z}) } e^{ik\cdot x(z, \bar{z})}
.}
For the vertex operator
to be on shell $-k^2 = M^2$ where $M^2$ is given in \mass . 
Similarly  the vertex operator for the complex conjugate tachyon is
given by
\eqn\tvertexa{
\bar{{\cal T}} =  e^{-\phi(z)} e^{-\tilde{\phi}(\bar{z})}
\sigma_{-\nu}(z, \bar{z}) e^{-i (\nu-1)H(z)} e^{+i(\nu
-1)\tilde{H}(\bar{z}) }
e^{-i\frac{wR}{2} (Y(z) - \tilde{Y}(\bar{z}) } e^{ik\cdot x(z,
\bar{z}) }
.}
To compute the four point amplitude we need to find the tachyon
vertex operator in the $(0,0)$ picture. This is done by the action of
the picture changing operator 
$e^{\phi} T_F e^{\tilde{\phi} } \tilde{T}_F$  on 
the operators in \tvertex\ and \tvertexa. 
$T_F$ is the supercharge given by
\eqn\supercharge{
T_F = -\frac{1}{4} ( \partial X^+ \psi^- +  \partial X^- \psi^+ )
-\frac{1}{2} ( \partial Y \psi^Y + \partial x^\mu \psi_\mu)
.}
Here $\psi^Y$ and $\psi^\mu$ are superpartners of $Y$ and $x^\mu$
respectively.

The four point scattering amplitude is given by
\eqn\fpt{
{\cal A} = 2\pi \int d^2x
\langle
c(\infty) \tilde{c}(\infty) 
\bar{{\cal T}} (\infty , \infty ) 
c(1) \tilde{c}(1)
e^\phi T_F e^{\bar{\phi} } \bar{T}_F {\cal T} (1)
\bar{{\cal T}} (x, \bar{x} ) 
c(0) \tilde{c}(0)
e^\phi T_F e^{\bar{\phi} } \bar{T}_F {\cal T} (0)
\rangle.
}
One can show that the excited twist operators do not contribute in this
correlation function. Let $k_4, k_3, k_2, k_1$ be the space time
momentum of the tachyon operators located at $\infty, 1, x, 0 $
respectively.
Evaluating all the correlation functions we get
\eqn\forpt{
{\cal A} = 2\pi \int d^2x 
\left(k_3\cdot k_1 + \frac{w^2R^2}{4} \right)^2
|1-x|^{2k_3\cdot k_2 -4h} |x|^{2k_2\cdot k_1 -4h} I(x, \bar{x} )
,}
where
\eqn\defh{
h= \frac{1-\nu}{2} + \frac{w^2R^2}{8}
}
and $I(x, \bar{x})$ is given by
\eqn\defi{
I(x, \bar{x}) = (F(x)F(1-\bar{x}) + F(\bar{x})F(1-x))^{-1}
,}
where $F(x)$ is the hypergeometric function $F(\nu, 1-\nu, 1; x)$
\foot{There is no lattice sum involved as the directions $X^+$ and
$X^{-}$ are non-compact.}.

\subsec{Cubic couplings of the tachyon}
We find the cubic couplings of the tachyon by factorizing the
four-point amplitude into its various channels.
To extract the $s = (k_1 + k_2)^2$, 
channel process look at the limit $x\rightarrow 0$.
Using the following property
\eqn\ilimit{
\lim_{x\rightarrow0} I(x, \bar{x})  = 1
}
and integrating around $x=0$ in \forpt
we get
\eqn\schan{
{\cal A}
= 4\pi^2  \left(k_3\cdot k_1 + \frac{w^2 R^2}{4} \right)^2 \frac{1}{s}
.}
Similarly the amplitude factorized along the $t= (k_3+ k_2)^2$, 
channel is given by
\eqn\tchan{
{\cal A }
= 4\pi^2 \left( k_3 \cdot k_1 + \frac{w^2R^2}{4} \right)^2 \frac{1}{t}
.}

Both these terms arise from exchange of the graviton, the
antisymmetric tensor, the dilaton and the scalars that resulted from
the compactification of the graviton and the antisymmetric tensor. 
From \schan\ and \tchan\ the cubic couplings of the tachyon with the
massless fields can be read out. The zero momentum tachyon couples
only to the fluctuation of the $g_{yy}$ component of the metric. 
Therefore the cubic coupling 
is given by
\eqn\cubcoup{
2\pi \int d^7x \frac{w^2R^2}{4} T\bar{T} \varphi
,}
here $T$ stands for the tachyon and $\varphi$ is the fluctuation of the
$g_{yy}$ component of the metric.
Note that 
the OPE of a tachyon and its complex conjugate results only in the
untwisted sector due to  charge conservation.
Restricting the factorization only to the fields which
survive the zero slope limit we
obtain only the massless fields discussed above.

It is possible that there might be cubic 
couplings of two tachyons with $w=1$ to a state of winding $w=2$ with
mass given by \mass. We show that this is not possible for $2\nu <1$,
which holds in the zero slope limit. 
We see that 
two tachyons
with $w=1$ couple only to states with masses of the order of string
scale. 

The cubic couplings of two tachyons with $w=1$ to a state with $w=2$
can be extracted from the $u= (k_2 + k_4)^2$, channel.
We can extract the leading exchange in the $u$ channel
by taking the limit $x\rightarrow \infty$. The following identities
are helpful in obtaining this limit.
\eqn\identity{\eqalign{
F(\nu, 1-\nu, 1; x) = (1-x)^{-\nu} F(\nu, \nu, 1, \frac{x}{x-1}),
 \cr
F(\nu, \nu, 1; 1) = \frac{\Gamma (1-2\nu)}{\Gamma^2(1-\nu))}
\;\;\;\;\;\; (1-2\nu >0)
.}}
The leading $u-$channel exchange in the four-point amplitude is given by
\eqn\uchan{
4\pi^2 \left(k_3\cdot k_1 + \frac{w^2 R^2}{4}\right) ^2
\frac{\Gamma^4(1-\nu) }{2\cos\pi\nu (\Gamma^2 (1-2\nu) ) } 
\frac{1}{u + 2 -6\nu + w^2R^2}
.}
Note that the leading 
pole is due to the exchange of a particle of string
mass. This state can be created by the following vertex operator in
the $(0, 0)$ picture.
\eqn\ustate{
\int d^2z \sigma_{2\nu}(z, \bar{z})  e^{2i(\nu-1)H(z)}
e^{-2i(\nu-1)\tilde{H} (\bar{z}) } 
e^{wR(Y(z) -\tilde{Y}(\bar{z}) ) }
.}
Thus we conclude that the zero momentum tachyon has cubic couplings
only with the metric fluctuation of $g_{yy}$, $\varphi$ in the zero slope
limit.

\subsec{Quartic coupling of the tachyon}

We compute the quartic coupling of the tachyon by subtracting the $s$
channel and $t$ channel poles from the four-point amplitude. 
As we are interested in the tachyon potential in the zero slope limit
it is consistent to set $\nu=0$ to obtain the leading term in the
four-point amplitude. We are also expanding the tachyon potential when
the tachyon is almost marginal, thus all the momenta can be set to
zero in the on-shell amplitude.
Setting $\nu=0$ using marginality of the tachyon in \forpt\ we obtain
\eqn\forptmar{
{\cal A} = 2\pi^2
\left( k_3\cdot k_1 + \frac{w^2R^2}{4} \right) ^2 \frac{\Gamma(t/2)
\Gamma(s/2) \Gamma (u/2 +1) }{\Gamma(-u/2) \Gamma(1-s/2) \Gamma(1-t/2)
}. }
Subtracting the $s$ and $t$ channel poles and setting the momenta to
zero we obtain  
\eqn\quartic{
-16\pi^2 \gamma \frac{w^4R^4}{16}
,}
where $\gamma= .5772$, the
Euler-Mascheroni constant. Thus the zero momentum quartic coupling for
a marginal tachyon is given by
\eqn\qcoup{
-16\pi^2 \gamma \int d^7x \frac{w^4R^4}{16} (T\bar{T})^2
.}

\appendix{E}{RG flows near the critical radius}

As we have discussed in section 4.3, the fluxbrane ceases to be unstable
upon compactifying $y$ (see \background) on a circle of radius $R'<R'_{\rm cr} = 1/2q$ 
as its lowest-momentum tachyon mode is actually massive. 
Consequently, the fluxbrane undergoes a `phase transition' at $R=R'_{\rm cr}$.

The tachyon at $R$ marginally greater than
$R'_{\rm cr}$ is almost massless. It is of interest to ask: Does its condensation
an endpoint to tachyon condensation `near' the inital unstable flux brane?
Relatedly, just below the critical radius, is the flux brane stable to small
but finite perturbations. The picture developed in section 3 of this paper
would suggest the answers in the negative to these questions. 
We independently confirm these answers in this Appendix. Our analysis 
uses the results derived in Appendix D.

In the neighborhood of the critical radius, and for the 
small values of the tachyon, 
it is possible to argue (see Appendix D) 
that tachyon dynamics is governed 
by an effective potential that 
contains only two modes; the zero momentum
tachyon $T$ and $\varphi$, the zero momentum mode corresponding to the
fluctuation of the metric component  $g_{yy}$. 
It is a modulus of the theory.
Physically this mode
corresponds to change in the radius $R'$ and the twist parametrized by
$q$. This change is given by\foot{The factor $4\pi$ occurs in relating the fluctuation 
coefficient in the vertex operator to the fluctuations of the metric,
see equation (6.6.18) of \PolchinskiRQ. }
\eqn\change{\eqalign{
R' &\rightarrow \frac{R'}{\sqrt{1+4\pi\varphi}}, \cr
q &\rightarrow \frac{q}{\sqrt{1+4\pi\varphi}}
.}}
It is easy to see that the total flux which is proportional to the
ratio $q/R'$ does not change under a change in $\varphi$, 
reflecting the 
fact that $\varphi$ is a modulus.
From the three-point and the four-point couplings of the tachyon we
obtain the following potential
\eqn\tacpot{
V = 
-\frac{1}{2}
\left(2qk - k^2 \right)
T\bar{T} + 2\pi k^2 T\bar{T} \varphi
-16\pi^2 \gamma k^4 (T\bar{T})^2
}
($\gamma$ is the Euler-Mascheroni constant). Here $k$ is the momentum of
the tachyon in the $y$ direction, and it is assumed that the tachyon is
almost marginal, i.e. $k={1 \over R'} \approx 2 q$. 
The coefficients of this
potential are computed in Appendix D using conformal field theory
techniques. 
From this potential it is easy to evaluate 
the beta functions for 
 $|T|$ and for $\varphi$. They are given by
\eqn\cftrg{\eqalign{
\dot{|T|} &= (2qk - k^2) |T|
- 4\pi k^2 |T|\varphi +
64\pi^2\gamma k^4 |T|^3, \cr
\dot{\varphi}&= -2\pi k^2 |T|^2.
}}
The phase portraits for these flows are plotted below. 
Note that this phase diagram has no nontrivial fixed points 
(i.e. fixed points with $|T| \neq 0$) 
irrespective of whether the tachyon
mass is positive or negative. In fact the only fixed points in this
approximation are on the line $|T|=0$. 
These fixed points are stable for 
$\varphi>\varphi^*$ where 
\eqn\cirtcial{
\varphi^* = \frac{2q-k}{4\pi k}.
}
Thus $\varphi^*$ is the value of of $\varphi$ at which the linear
effective mass of the tachyon in the first equation of \cftrg\
vanishes. The separatrix ABC, separates flows that end on 
line $|T|=0$  for $\varphi>\varphi^*$ and flows that runaway to
regions beyond the approximation of small $|T|$.
Notice that a slight perturbation away from the fixed point
$\varphi=0$, $|T|=0$ takes the 
flow far away, beyond the approximation of
small $|T|$. 
\fig{Sketch of RG flow in the $|T|$-$\varphi$ plane.
The trajectory ABC is the separatrix.}{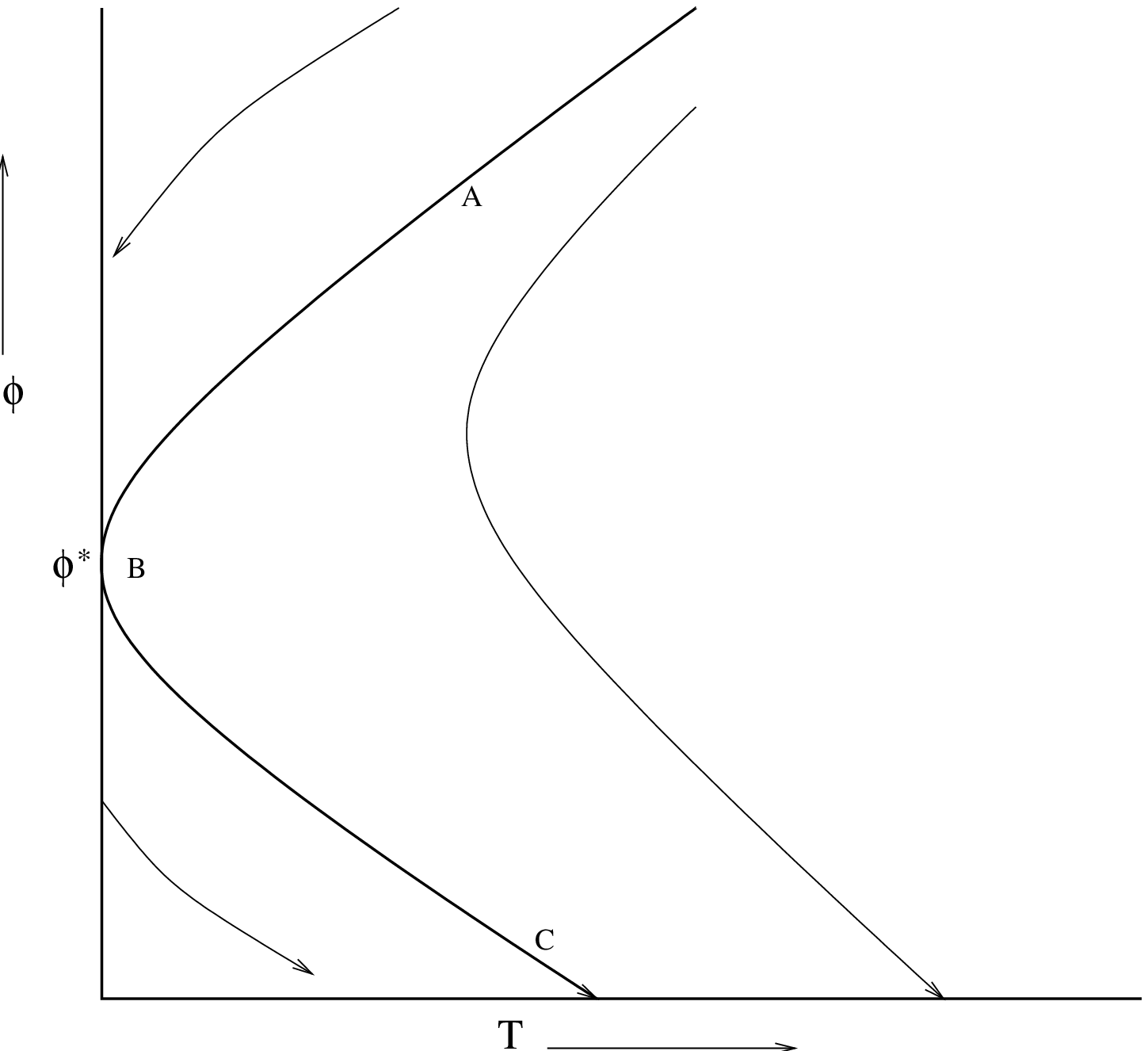}{3.5truein}

We were able to qualitatively confirm 
these results with our numerical simulations.
which clearly indicated that 
just below the critical radius the system was
not stable against finite perturbations.

\listrefs

\end